\documentclass[preprint2]{aastex631}

\shorttitle{JWST NIRISS - V. KPI}
\shortauthors{Kammerer et al.}

\usepackage{amsmath}
\usepackage{amssymb}
\usepackage{bm}
\usepackage{graphicx}
\usepackage{makecell}
\begin{document}

\title{The Near Infrared Imager and Slitless Spectrograph for \emph{JWST} -\\V. Kernel Phase Imaging and Data Analysis}

\correspondingauthor{Jens Kammerer}
\email{jkammerer@stsci.edu}

\author[0000-0003-2769-0438]{Jens Kammerer}
\affiliation{Space Telescope Science Institute, 3700 San Martin Dr, Baltimore, MD 21218, USA}

\author[0000-0001-7864-308X]{Rachel A. Cooper}
\affiliation{Space Telescope Science Institute, 3700 San Martin Dr, Baltimore, MD 21218, USA}

\author[0000-0002-5922-8267]{Thomas Vandal}
\affiliation{Institut de Recherche sur les Exoplan\`etes (iREx), Universit\'e de Montr\'eal, D\'epartement de Physique, C.P. 6128 Succ. Centre-ville, Montr\'eal, QC H3C 3J7, Canada}

\author{Deepashri Thatte}
\affiliation{Space Telescope Science Institute, 3700 San Martin Dr, Baltimore, MD 21218, USA}

\author[0000-0003-1180-4138]{Frantz Martinache}
\affiliation{Universit\'e C\^ote d’Azur, Observatoire de la C\^ote d’Azur, CNRS, Laboratoire Lagrange, France}

\author[0000-0003-1251-4124]{Anand Sivaramakrishnan}
\affiliation{Space Telescope Science Institute, 3700 San Martin Dr, Baltimore, MD 21218, USA}
\affiliation{Astrophysics Department, American Museum of Natural History, 79th Street at Central Park West, New York, NY 10024, USA}
\affiliation{Department of Physics and Astronomy, Johns Hopkins University, 3701 San Martin Dr, Baltimore, MD 21218, USA}

\author{Alexander Chaushev}
\affiliation{Department of Physics and Astronomy, 4129 Frederick Reines Hall, University of California, Irvine, CA 92697, USA}

\author[0000-0002-5823-3072]{Tomas Stolker}
\affiliation{Leiden Observatory, Leiden University, Niels Bohrweg 2, 2333 CA Leiden, The Netherlands}

\author{James P. Lloyd}
\affiliation{Carl Sagan Institute/Department of Astronomy and Space Sciences, Cornell University,
Ithaca, NY 14853, USA}

\author[0000-0003-0475-9375]{Lo\"ic Albert}
\affiliation{Institut de Recherche sur les Exoplan\`etes (iREx), Universit\'e de Montr\'eal, D\'epartement de Physique, C.P. 6128 Succ. Centre-ville, Montr\'eal, QC H3C 3J7, Canada}

\author{Ren\'e Doyon}
\affiliation{Institut de Recherche sur les Exoplan\`etes (iREx), Universit\'e de Montr\'eal, D\'epartement de Physique, C.P. 6128 Succ. Centre-ville, Montr\'eal, QC H3C 3J7, Canada}
\affiliation{Observatoire du Mont-M\'egantic, Universit\'e de Montr\'eal, Montr\'eal, QC H3C 3J7, Canada}

\author{Steph Sallum}
\affiliation{Department of Physics and Astronomy, 4129 Frederick Reines Hall, University of California, Irvine, CA 92697, USA}

\author{Marshall D. Perrin}
\affiliation{Space Telescope Science Institute, 3700 San Martin Dr, Baltimore, MD 21218, USA}

\author{Laurent Pueyo}
\affiliation{Space Telescope Science Institute, 3700 San Martin Dr, Baltimore, MD 21218, USA}
\affiliation{Department of Physics \& Astronomy, Johns Hopkins University, 3400 N. Charles Street, Baltimore, MD 21218, USA}

\author{Antoine M\'erand}
\affiliation{European Southern Observatory, Karl-Schwarzschild-Straße 2, 85748 Garching, Germany}

\author[0000-0001-7853-4094]{Alexandre Gallenne}
\affiliation{Universidad de Concepci\'on, Departamento de Astronom\'ia, Casilla 160-C, Concepci\'on, Chile}
\affiliation{Unidad Mixta Internacional Franco-Chilena de Astronom\'ia (CNRS UMI 3386), Departamento de Astronom\'ia, Universidad de Chile, Camino El Observatorio 1515, Las Condes, Santiago, Chile}

\author[0000-0002-7162-8036]{Alexandra Greenbaum}
\affiliation{IPAC, Mail Code 100-22, Caltech, 1200 E. California Blvd, Pasadena, CA 91125, USA}

\author{Joel Sanchez-Bermudez}
\affiliation{Instituto de Astronom\'ia, Universidad Nacional Aut\'onoma de M\'exico, Apdo. Postal 70264, Ciudad de M\'exico, 04510, Mexico}
\affiliation{Max-Planck-Institut f\"ur Astronomie, K\"onigstuhl 17, D-69117 Heidelberg, Germany}

\author[0000-0001-9582-4261]{Dori Blakely}
\affiliation{Department of Physics and Astronomy, University of Victoria, Victoria, BC V8P 5C2, Canada}
\affiliation{NRC Herzberg Astronomy and Astrophysics, 5071 West Saanich Rd, Victoria, BC V9E 2E7, Canada}

\author[0000-0002-6773-459X]{Doug Johnstone}
\affiliation{NRC Herzberg Astronomy and Astrophysics, 5071 West Saanich Rd, Victoria, BC V9E 2E7, Canada}
\affiliation{Department of Physics and Astronomy, University of Victoria, Victoria, BC V8P 5C2, Canada}

\author[0000-0002-3824-8832]{Kevin Volk}
\affiliation{Space Telescope Science Institute, 3700 San Martin Dr, Baltimore, MD 21218, USA}

\author{Andre Martel}
\affiliation{Space Telescope Science Institute, 3700 San Martin Dr, Baltimore, MD 21218, USA}

\author[0000-0002-5728-1427]{Paul Goudfrooij}
\affiliation{Space Telescope Science Institute, 3700 San Martin Dr, Baltimore, MD 21218, USA}

\author[0000-0003-1227-3084]{Michael R. Meyer}
\affiliation{Astronomy Department, University of Michigan, Ann Arbor, MI 48109, USA}

\author[0000-0002-4201-7367]{Chris J. Willott}
\affiliation{NRC Herzberg Astronomy and Astrophysics, 5071 West Saanich Rd, Victoria, BC V9E 2E7, Canada}

\author[0000-0003-1863-4960]{Matthew De Furio}
\affiliation{Department of Astronomy, University of Michigan, Ann Arbor, MI 48109, USA}

\author[0000-0003-4987-6591]{Lisa Dang}
\affiliation{Department of Physics, McGill University, 3600 Rue University, Montréal, QC H3A 2T8, Canada}
\affiliation{Institut de Recherche sur les Exoplan\`etes (iREx), Universit\'e de Montr\'eal, D\'epartement de Physique, \\C.P. 6128 Succ. Centre-ville, Montr\'eal, QC H3C 3J7, Canada}

\author[0000-0002-3328-1203]{Michael Radica}
\affiliation{Institut de Recherche sur les Exoplan\`etes (iREx), Universit\'e de Montr\'eal, D\'epartement de Physique, C.P. 6128 Succ. Centre-ville, Montr\'eal, QC H3C 3J7, Canada}

\author{Ga\"el Noirot}
\affiliation{Department of Astronomy \& Physics and Institute for Computational Astrophysics, Saint Mary’s University, 923 Robie St, Halifax, NS B3H 3C3, Canada}

%%%%%%%%%%%%%%%%%%%%%%%

\begin{abstract}
%-CONTEXT
Kernel phase imaging (KPI) enables the direct detection of substellar companions and circumstellar dust close to and below the classical (Rayleigh) diffraction limit. The high-Strehl full pupil images provided by the \emph{James Webb Space Telescope} (\emph{JWST}) are ideal for application of the KPI technique.
%-AIMS
We present a kernel phase analysis of \emph{JWST} NIRISS full pupil images taken during the instrument commissioning and compare the performance to closely related NIRISS aperture masking interferometry (AMI) observations.
%-METHODS
For this purpose, we develop and make publicly available the custom \texttt{Kpi3Pipeline} data reduction pipeline enabling the extraction of kernel phase observables from \emph{JWST} images. The extracted observables are saved into a new and versatile kernel phase FITS file (KPFITS) data exchange format. Furthermore, we present our new and publicly available \texttt{fouriever} toolkit which can be used to search for companions and derive detection limits from KPI, AMI, and long-baseline interferometry observations while accounting for correlated uncertainties in the model fitting process.
%-RESULTS
Among the four KPI targets that were observed during NIRISS instrument commissioning, we discover a low-contrast ($\sim1$:5) close-in ($\sim1~\lambda/D$) companion candidate around CPD-66~562 and a new high-contrast ($\sim1$:170) detection separated by $\sim1.5~\lambda/D$ from 2MASS~J062802.01-663738.0. The 5--$\sigma$ companion detection limits around the other two targets reach $\sim6.5$~mag at $\sim200$~mas and $\sim7$~mag at $\sim400$~mas. Comparing these limits to those obtained from the NIRISS AMI commissioning observations, we find that KPI and AMI perform similar in the same amount of observing time.
%-CONCLUSIONS
Due to its 5.6~times higher throughput if compared to AMI, KPI is beneficial for observing faint targets and superior to AMI at separations $\gtrsim325$~mas. At very small separations ($\lesssim100$~mas) and between $\sim250$--325~mas, AMI slightly outperforms KPI which suffers from increased photon noise from the core and the first Airy ring of the point-spread function.

\end{abstract}

\section{Introduction}
\label{sec:introduction}

The recently commissioned \emph{James Webb Space Telescope} (\emph{JWST}) is a joint NASA/ESA/CSA flagship science mission to explore the beginnings of the Universe, the assembly of galaxies, the birthplaces of stars, and planetary systems and the origins of life \citep{gardner2006}. For the direct observation and characterization of giant exoplanets and circumstellar dust, \emph{JWST} is equipped with multiple coronagraphic imaging modes in the near- and mid-infrared as well as a non-redundant mask (NRM) operating between $\sim2.8$--$4.8~\text{\textmu m}$ that transforms the 6.5~m primary mirror into an interferometric array of seven subapertures \citep{artigau2014}. It is the first time in history that such an aperture mask is available on a space-based telescope. By exploiting the interferometric capabilities of this mask, the NIRISS instrument \citep{doyon2012} is expected to enable high-contrast imaging up to $\sim10$~mag contrast at small angular separations of $\sim70$--400~mas \citep{greenbaum2015}, well inside the inner working angle (IWA) of the NIRCam coronagraphs \citep[half width half maximum of $\sim6~\lambda/D$, where $\lambda$ is the observing wavelength and $D$ is the telescope primary mirror diameter,][]{krist2009,krist2010}. The unique parameter space accessible with the NIRISS NRM enables advances in exoplanet and planet formation as well as galaxy evolution science by studying close-in stellar and substellar companions, warm exozodiacal dust around nearby stars, transitional disks, and feedback in active galactic nuclei for instance \citep{sivaramakrishnan2022}.

While the NIRISS NRM transforms \emph{JWST} into an interferometer and thereby enables high-contrast imaging down to the Michelson diffraction limit ($\sim0.5~\lambda/D$), the NRM also blocks $\sim85\%$ of the incoming light and thus significantly reduces the sensitivity of the observations \citep{artigau2014}. An alternative technique that is not affected by such a big throughput loss is called ``kernel phase imaging'' (KPI) and has been introduced by \citet{martinache2010}. KPI uses Fourier quantities of full pupil images to achieve the same high resolution ($\sim0.5~\lambda/D$) as aperture masking interferometry (AMI). The basic idea of KPI is to discretize the full pupil into an interferometric array of ``virtual'' subapertures during post-processing. However, due to the high redundancy of the full pupil, KPI requires high-Strehl images (which are naturally given from space in the absence of a disturbing atmosphere) so that the image Fourier phase can be linearized and the contributions from individual baselines can be disentangled. In this linear regime, the relationship between the image Fourier phase $\phi$ and the pupil plane phase $\varphi$ reads
\begin{equation}
    \phi = \bm{R}^{-1}\cdot\bm{A}\cdot\varphi+\phi_\text{obj},
    \label{eqn:fourier_phase}
\end{equation}
where $\bm{A}$ is the matrix mapping baselines in the pupil plane to spatial frequencies in the image Fourier plane, $\bm{R}$ is the diagonal matrix encoding the redundancy of each baseline, and $\phi_\text{obj}$ is the image Fourier phase intrinsic to the observed object. One can then multiply Equation~\ref{eqn:fourier_phase} with $\bm{R}$ from the left, obtain the kernel $\bm{K}$ of the baseline mapping matrix $\bm{A}$ via singular value decomposition, and derive the kernel phase
\begin{align}
    \theta &= \bm{K}\cdot\bm{R}\cdot\phi\\
    &= \bm{K}\cdot\bm{A}\cdot\varphi+\bm{K}\cdot\bm{R}\cdot\phi_\text{obj}\\
    &= \bm{K}\cdot\bm{R}\cdot\phi_\text{obj}\\
    &= \theta_\text{obj}
\end{align}
which is independent of pupil plane phase errors to first order \citep[higher order terms appear because the linearization applied here is only an approximation,][]{ireland2013}. Therefore, kernel phase has similar properties as closure phase in AMI and \citet{ireland2016} has shown that kernel phase is a generalization of closure phase to the case of redundant apertures. Hence, KPI with \emph{JWST} is expected to achieve similar performance as AMI except for an increased sensitivity to faint targets due to its increased throughput.

With NIRISS, KPI can be used with the same four filters as AMI, which are F277W, F380M, F430M, and F480M, albeit the usefulness of F277W is limited given that the NIRISS detector is undersampled at this wavelength. The AMI observing template, which is also used for KPI observations, provides repeatable target acquisition on a predefined detector position to subpixel accuracy \citep[typically $<0.1$~pixels][]{rigby2022}. Using the same detector position for the science and the PSF reference target will help mitigating systematic errors (e.g., from flat-fielding). The methods presented here are based on a monochromatic description of the image and given the $\sim6\%$ bandwidth of the medium-band filters mentioned above, we expect some level of spectral decoherence for both AMI and KPI techniques. While a detailed study of this effect is beyond the scope of this paper, we note that kernel phase techniques have been successfully applied to \emph{Hubble Space Telescope} (\emph{HST}) wide-band (F110W) images by \citet{pope2013}. To minimize calibration errors, we recommend to use a PSF reference target with a similar spectral type as the science target. In the model fitting process with \texttt{fouriever} (Section~\ref{sec:model_fitting_with_fouriever}), it is possible to apply bandwidth smearing to account for the finite filter bandpass of the observations. Finally, since KPI can be applied to any full pupil image, it can be used with all instruments onboard \emph{JWST} opening up the entire wavelength range of $\sim0.7$--$25.5~\text{\textmu m}$ for high-resolution imaging. KPI's increased uv-coverage with respect to AMI also enables high-contrast image reconstruction of complex scenes down to the Michelson criterion. For a more detailed description of kernel phase we refer the reader to \citet{martinache2010} and \citet{martinache2020}.

The first application of KPI can be found in \citet{martinache2010} who detected a previously known companion to the star GJ~164 at a contrast of $\sim9:1$ and a separation of $\sim0.6~\lambda/D$ in archival \emph{HST} NICMOS data, clearly demonstrating the super-resolution power of KPI. Later, \citet{pope2013} discovered five new brown dwarf companions with separations ranging from $\sim0.4$--$0.6~\lambda/D$, also in archival \emph{HST} NICMOS data. The first time that KPI was applied behind an adaptive optics system from the ground was in \citet{pope2016} who observed the known binary $\alpha$~Oph with both KPI and AMI and found excellent agreement between the binary parameters derived from both techniques. \citet{kammerer2019} and \citet{wallace2020} went one step further and searched for close-in substellar companions in an archival VLT NACO dataset of nearby field stars and a Keck NIRC2 dataset of young stars in the Taurus Molecular Cloud where they achieved contrast limits of up to $\sim7$~mag at $\sim1~\lambda/D$ in the L-band. \citet{kammerer2019} and \citet{laugier2019} came up with different methods to extend the dynamic range of KPI observations beyond the saturation limit in the central few pixels of the core of the point-spread function (PSF). The method from \citet{kammerer2019} is also implemented in the \emph{JWST} kernel phase pipeline presented here and explained in more detail in Section~\ref{sec:bad_pixel_fixing_step}. Similar to angular differential imaging \citep{marois2006} in classical high-contrast imaging, \citet{laugier2020} also derived self-calibrating angular differential kernel phase observables, but these are more suitable for datasets with larger field rotation than \emph{JWST} can provide. To measure the photometry of the individual components of the T~Tau triple star system at $\sim10~\text{\textmu m}$, \citet{kammerer2021} applied KPI to VLT VISIR-NEAR observations in the mid-infrared, demonstrating the feasibility of KPI with instruments such as \emph{JWST} MIRI. Recently, pioneering work from \citet{pope2021} has shown that automatic differentiation methods can be used to extract kernel phase observables from any differentiable optical system, such as a Lyot coronagraph for instance, which has previously been impossible given that focal plane masks destroy the linearity of the Fourier phase observables. Albeit beyond the scope of this paper, automatic differentiation methods are a viable alternative beyond the classical KPI methods presented here.

The paper is structured as follows. Section~\ref{sec:methods} describes the methods that we use to reduce and analyze the \emph{JWST} NIRISS KPI commissioning data. In particular, Section~\ref{sec:kernel_phase_stage_3_pipeline} introduces our customly developed and publicly available stage 3 kernel phase data reduction pipeline that can be used to extract kernel phase observables from \emph{JWST} images in a similar fashion as the other STScI \texttt{jwst}\footnote{\url{https://github.com/spacetelescope/jwst}} data reduction pipelines. Section~\ref{sec:kernel_phase_fits_files} motivates the need for a dedicated kernel phase data exchange format and describes the KPFITS file format that we are proposing here. Section~\ref{sec:model_fitting_with_fouriever} introduces our publicly available model fitting toolkit \texttt{fouriever} that can be used to search for companions or determine detection limits from OIFITS \citep[a widely used file format for exchanging AMI and long-baseline interferometry data,][]{pauls2005,duvert2017} and KPFITS files. The NIRISS KPI commissioning observations are described in Section~\ref{sec:observations} and our results from their analysis are presented and discussed in the context of the NIRISS AMI commissioning observations in Section~\ref{sec:results_and_discussion}. Finally, Section~\ref{sec:summary_and_conclusions} summarizes our work and draws conclusions for the application of AMI and KPI with \emph{JWST}.

\section{Methods}
\label{sec:methods}

For reducing and analyzing the \emph{JWST} NIRISS KPI commissioning data, we develop a variety of custom and user-friendly software that we make publicly available on GitHub. This enables other \emph{JWST} observers to use our tools for proposal planning and for getting science-ready data products and plots out of their own programs.

\subsection{Kernel phase stage 3 pipeline}
\label{sec:kernel_phase_stage_3_pipeline}

The STScI \texttt{jwst} data reduction pipeline is organized into three stages. Stage 1 performs detector level calibrations such as bias correction and jump detection, stage 2 performs photometric calibrations, and stage 3 performs high-level calibrations that are specific to each of the observing modes supported by \emph{JWST} (e.g., AMI, coronagraphic imaging, integral field spectroscopy). To enable the community a straightforward access to kernel phase observables from \emph{JWST} images, we develop a custom stage 3 pipeline that we name the \texttt{Kpi3Pipeline}\footnote{https://github.com/kammerje/jwst-kpi}. This pipeline can be used for extracting kernel phase observables from full pupil NIRISS and NIRCam images, support for MIRI will be added in a future update. Before the \emph{JWST} images can be fed into the \texttt{Kpi3Pipeline}, they need to be processed with the \texttt{jwst} stage 1 and 2 pipelines. For those, similar as with AMI data, we recommend to skip the inter-pixel capacitance, the photometry, and the resample step since Fourier plane imaging techniques in general are highly sensitive to variations in the pixel-by-pixel response of the detector \citep[e.g.,][]{ireland2013}. Skipping these steps avoids systematic errors being introduced by the pipeline corrections. Instead, the approach with NIRISS KPI and AMI observations is to center the science and the reference targets on the exact same detector pixel and thereby minimize flat-fielding errors. The available detector positions were carefully chosen to provide both a good uv-coverage of the pupil support (which is challenging with NIRISS given its coarse sampling) and a clean detector region with only few bad pixels \citep{sivaramakrishnan2022}. Dithering is hence not recommended for NIRISS KPI (and AMI) observations. While the backend of the \texttt{Kpi3Pipeline} is based on \texttt{XARA}\footnote{\url{https://github.com/fmartinache/xara}} \citep{martinache2010,martinache2013,martinache2020}, the frontend is designed similarly to the STScI \texttt{jwst} pipelines to provide a uniform experience for the user. Similar to the VLT NACO kernel phase data reduction pipeline developed by \citet{kammerer2019}, the \texttt{Kpi3Pipeline} consists of five steps:
\begin{enumerate}
    \item bad pixel fixing step,
    \item recentering step,
    \item windowing step,
    \item kernel phase extraction step,
    \item empirical uncertainties step.
\end{enumerate}
An intermediate data product can be saved after each step so that the individual steps can be run separately and each of the five steps can generate a diagnostic plot for validation purposes. Each step and its diagnostic plot is described in more detail below. A list of the tunable parameters of each step can be found in Appendix~\ref{sec:kernel_phase_stage_3_pipeline_parameters}. In principle, all steps except for the fourth one can be skipped although it is highly recommended to run at least steps 1, 2, and 4 to ensure that the kernel phase extraction step is provided with properly preprocessed data.

\begin{figure*}[t!]
\centering
\includegraphics[trim={7cm 7cm 0cm 0cm},clip,width=\textwidth]{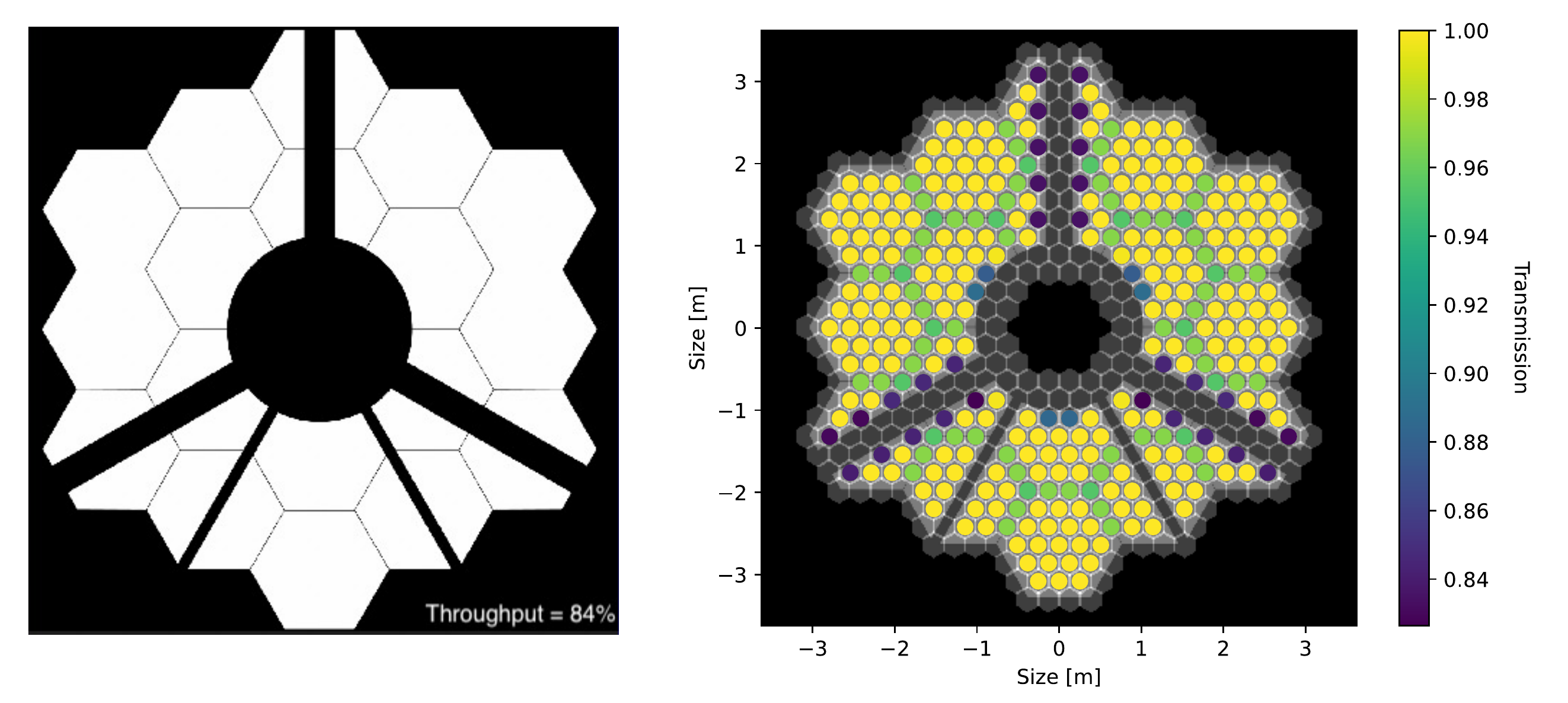}
\caption{Left: NIRISS CLEARP pupil mask used for full pupil imaging. The total throughput of the mask is $\sim84\%$. Right: pupil model used for the kernel phase analysis of the NIRISS commissioning data. The model consists of 349 individual subapertures spanning 948 distinct baselines. Each subaperture has a transmission $0.7 < T \leq 1$ that was determined by averaging the NIRISS CLEARP pupil mask transmission over the hexagonal grid shown in gray. Subapertures with a total transmission of $<0.7$ were discarded.}
\label{fig:pupil_model}
\end{figure*}

An additional input that is required for running the \texttt{Kpi3Pipeline} is a discrete model of the pupil. We note that due to different pupil plane masks, the NIRISS, NIRCam, and MIRI instruments all see slightly different pupils. We provide default pupil models for the NIRISS CLEARP and the NIRCam CLEAR pupils but the user can also provide a custom pupil model if desired. For the default pupil models, we distribute subapertures on a hexagonal grid so that 19 subapertures fall inside one \emph{JWST} primary mirror segment. To obtain an isotropic grid, we also distribute subapertures on top of the gaps between the individual primary mirror segments. We are using a ``gray'' pupil model so that the subapertures on top of the gaps have a slightly reduced transmission. Then, we apply the transmission of the instrument-specific pupil mask and discard all subapertures with a total transmission of $<0.7$. This results in the NIRISS CLEARP pupil model shown in Figure~\ref{fig:pupil_model}. This model consists of 349 individual subapertures spanning 948 distinct baselines. After discarding all baselines with a redundancy of less than 10, 800 distinct baselines yielding 452 individual kernel phases remain. Discarding baselines with low redundancy helps to avoid the longest (and most noisy) baselines at the edge of the primary mirror. We find that a redundancy threshold of 10 is sufficient to avoid Fourier phases $>0.5$~rad and entering the non-linear regime.

\subsubsection{Bad pixel fixing step}
\label{sec:bad_pixel_fixing_step}

\begin{figure*}[t!]
\centering
\includegraphics[trim={2cm 0cm 0.5cm 0cm},clip,width=\textwidth]{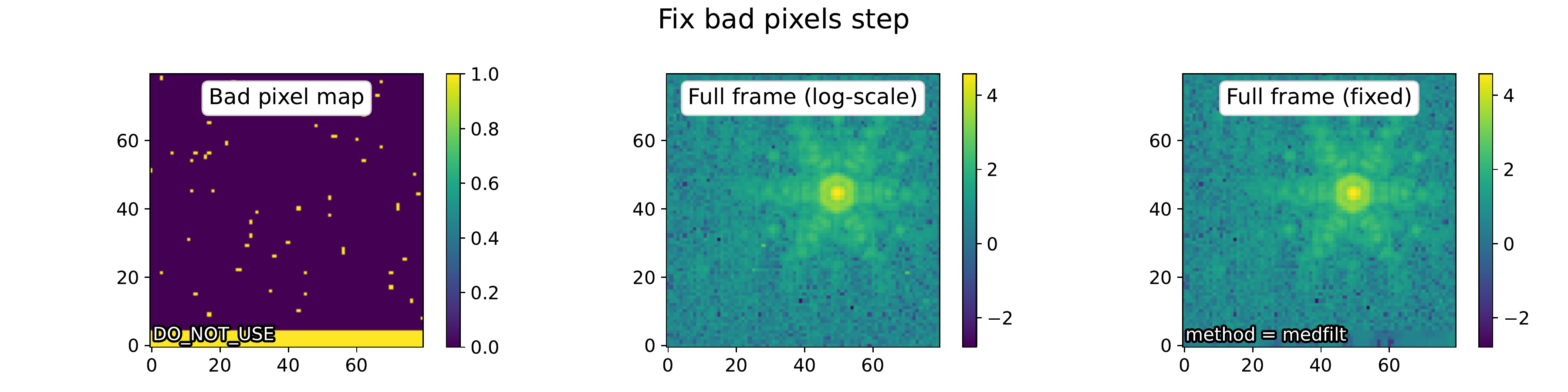}
\caption{Diagnostic plot produced by the bad pixel fixing step of the \texttt{Kpi3Pipeline} for the first frame of TYC~8906-1660-1 observed with \emph{JWST} NIRISS full pupil imaging. The left panel shows the bad pixel map from the \texttt{jwst} stage 1 pipeline considering the specified data quality flags as bad pixels (here DO\_NOT\_USE), the middle panel shows the frame in a logarithmic color stretch before fixing the bad pixels, and the right panel shows it after fixing the bad pixels using the ``medfilt'' method. The bottom five rows are reference pixels for correcting bias drifts which are not used for science.}
\label{fig:bad_pixel_fixing}
\end{figure*}

Bad pixels can have a ruinous impact on Fourier plane imaging since an individual bad pixel contributes noise to all spatial frequencies of the image. However, \citet{ireland2013} has shown that bad pixels do not affect kernel phase observables if they are properly corrected. In the \texttt{Kpi3Pipeline}, bad pixels can be fixed using two different methods: ``medfilt'' or ``fourier''. In both cases, the locations of the bad pixels are obtained from the \texttt{jwst} stage 1 pipeline. As a default, all pixels flagged as ``DO\_NOT\_USE'' in the data quality array are considered as bad pixels but the user may specify other data quality flags\footnote{\url{https://jwst-reffiles.stsci.edu/source/data_quality.html}} that shall be considered as bad pixels. With the ``medfilt'' method, bad pixels are simply replaced with the median filtered image using a kernel size of five pixels. With the ``fourier'' method, bad pixels are fixed by minimizing their Fourier power outside the region of support permitted by the pupil geometry. This method was introduced by \citet{ireland2013} and starts by computing the matrix $\bm{B}_Z$ which maps the bad pixel values $x$ onto the Fourier domain $Z$ (the complement of the pupil support), so that their Fourier power $\left|f_Z\right|$ is given by
\begin{equation}
    f_Z = \bm{B}_Z\cdot b+\epsilon_Z,
\end{equation}
where $b$ are the corrections to be made to the bad pixels and $\epsilon_Z$ is remaining noise. These corrections are obtained by computing the Moore-Penrose pseudo inverse of $\bm{B}_Z$ and solving for
\begin{equation}
    b = \bm{B}_Z^+\cdot f_Z = (\bm{B}_Z^*\cdot\bm{B}_Z)^{-1}\cdot\bm{B}_Z^*\cdot f_Z,
\end{equation}
where the star denotes the complex conjugate. This algorithm has been successfully applied to fix bad and reconstruct saturated pixels in kernel phase observations by \citet{kammerer2019} and is described in more detail in \citet{sivaramakrishnan2022}. While the ``fourier'' method is particularly suited for reconstructing saturated PSFs, we do not have any saturated PSFs in the NIRISS KPI commissioning data and therefore use the ``medfilt'' method for simplicity. The diagnostic plot of the bad pixel fixing step is shown in Figure~\ref{fig:bad_pixel_fixing}.

\subsubsection{Recentering step}

\begin{figure*}[t!]
\centering
\includegraphics[width=0.75\textwidth]{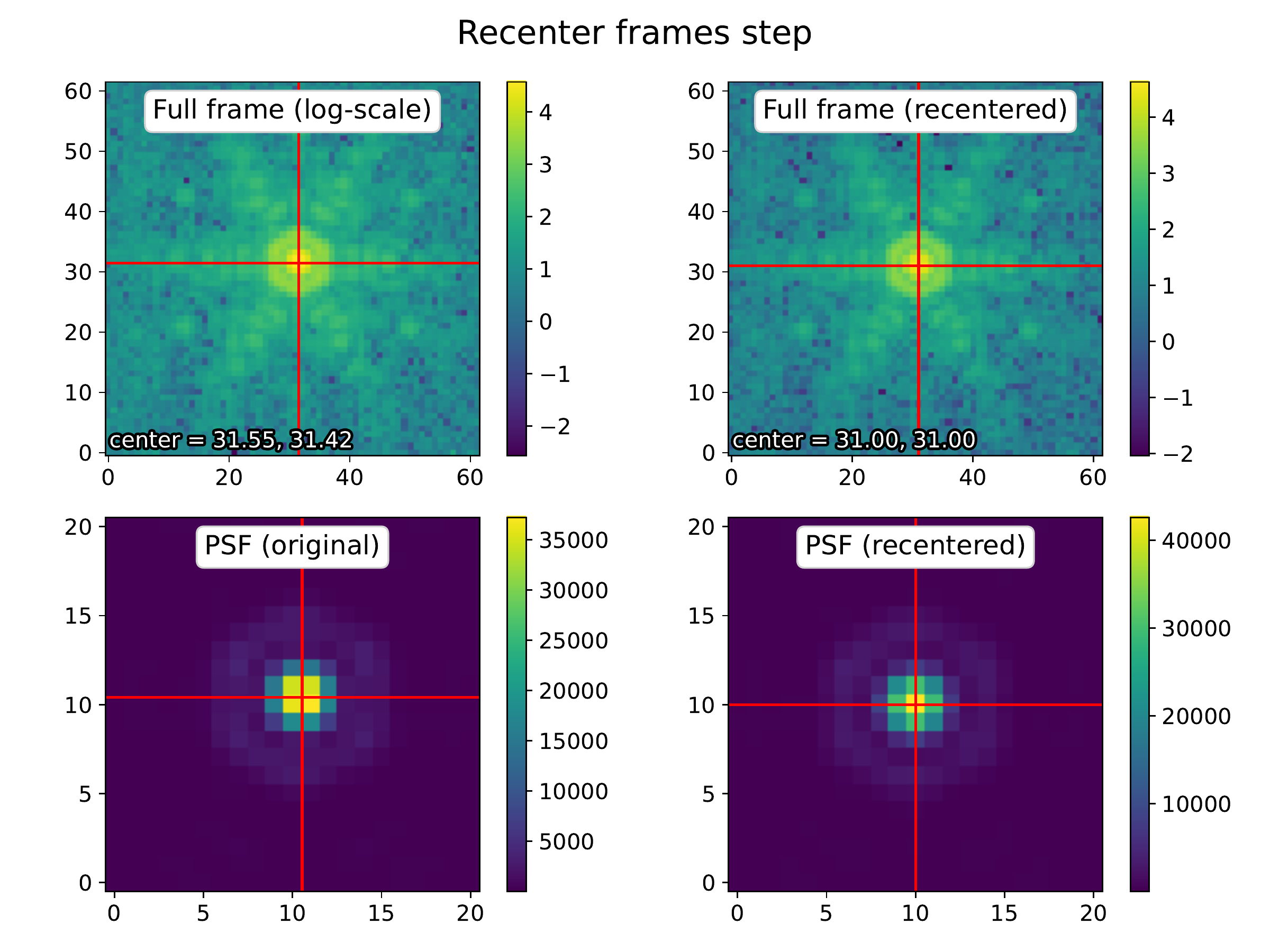}
\caption{Diagnostic plot produced by the recentering step of the \texttt{Kpi3Pipeline} for the same data as shown in Figure~\ref{fig:bad_pixel_fixing}. The left panels show the full frame (top) and a zoom on the PSF (bottom) before recentering and the right panels show them after recentering. The red crosshair indicates the identified PSF center.}
\label{fig:recentering}
\end{figure*}

As mentioned in Section~\ref{sec:introduction}, the image Fourier phase needs to be in the linear regime in order for the kernel phase technique to be applicable. Hence, recentering the images is important since a PSF being offset from the image center by only half of a pixel would already result in a linear ramp in the image Fourier phase from $-\pi/2$ to $+\pi/2$ across the support of the telescope pupil (i.e., the modulation transfer function of the pupil). While the kernel phase observables are independent of first order pupil plane phase errors \citep[and thus a linear ramp in the image Fourier phase,][]{martinache2010}, there are virtually always systematic phase errors that can easily cause the image Fourier phase wrapping around the $\pm\pi$ discontinuity if added on top of a linear phase ramp and thereby destroying the properties of the kernel phase observables. To avoid this, the \texttt{Kpi3Pipeline} uses the Fourier phase norm minimization (FPNM) method implemented in \texttt{XARA} that recenters the images using a least-squares gradient descent minimization of the norm of the image Fourier phase within the support of the telescope pupil. This method was shown to be more robust, especially in the presence of low-contrast companions at small angular separation \citep{kammerer2019}, albeit slightly slower than the other two methods (BCEN = centroid of the brightest speckle \& COGI = center of gravity of the image) that are also implemented in \texttt{XARA}. We note that before recentering the images, we trim them to their maximum possible square size (constrained by the location of the detector edges) around the expected PSF center (the reference pixel position where \emph{JWST} aims to place the PSF during target acquisition\footnote{This target acquisition is accurate to within $<0.1$~pixels \citep{rigby2022}.}). The diagnostic plot of the recentering step is shown in Figure~\ref{fig:recentering}.

\subsubsection{Windowing step}

\begin{figure*}[t!]
\centering
\includegraphics[width=0.8\textwidth]{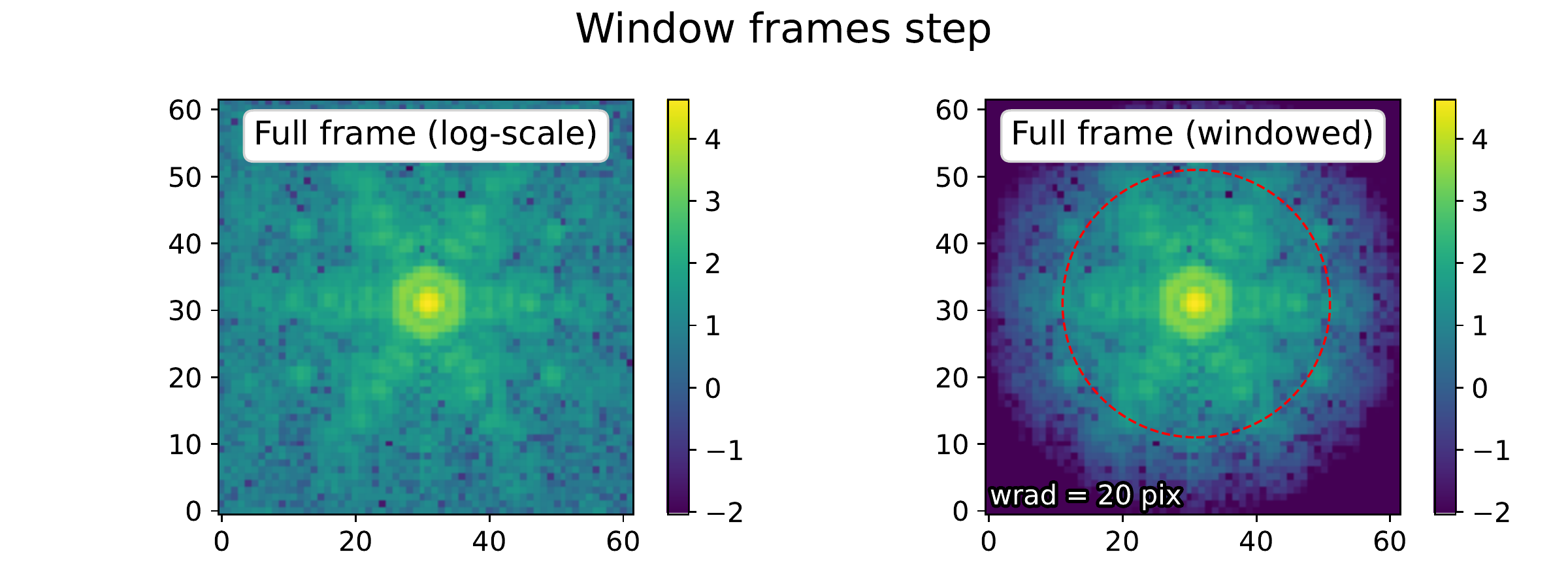}
\caption{Diagnostic plot produced by the windowing step of the \texttt{Kpi3Pipeline} for the same data as shown in Figure~\ref{fig:bad_pixel_fixing}. The left panel shows the trimmed and recentered image before windowing and the right panel shows it after windowing with a super-Gaussian windowing function with a radius of 20~pixels. For reference, the dashed red circle indicates a radius of 20~pixels from the center.}
\label{fig:windowing}
\end{figure*}

Windowing with a smooth function is an effective method to mitigate artifacts when numerically Fourier transforming images with sharp edges. Since \texttt{XARA} uses a Fourier transform to extract kernel phase observables from the telescope images, we window (i.e., multiply) them with a super-Gaussian windowing function $w$ of shape
\begin{equation}
    w(x,y) = \exp\left(-\frac{(x^2+y^2)^2}{r^4}\right)
\end{equation}
before Fourier transforming them, where $x$ and $y$ are the pixel coordinates with respect to the image center and $r$ is the radius of the super-Gaussian windowing function in units of pixels. The radius $r$ is chosen adaptively as the smaller of 40~pixels and a fourth of the trimmed square image size to make sure that the window is not larger than the image itself. If the images are large enough, a radius of 40~pixels gives access to separations of at least $5~\lambda/D$ (sometimes more depending on the observing wavelength). At these separations, classical high-contrast imaging techniques such as coronagraphy are typically superior to KPI. The NIRISS full pupil images considered for the KPI analysis here do only have a native size of 80 by 80~pixels and we are using a constant window radius of $r = 15$~pixels for them resulting in a field-of-view of $\sim2$~arcsec, which is more than sufficient for the region where NIRISS KPI is superior to NIRCam coronagraphy \citep{kammerer2022}. The diagnostic plot of the windowing step is shown in Figure~\ref{fig:windowing}.

\subsubsection{Kernel phase extraction step}

Once the bad pixels are fixed and the images are recentered and windowed, kernel phase observables can be extracted using \texttt{XARA}. First, the image Fourier phase $\phi$ at the discrete spatial frequencies of the pupil model is obtained using a linear discrete Fourier transform. Then, the kernel phase $\theta$ is computed as
\begin{equation}
    \theta = \Tilde{\bm{K}}\cdot\phi,
\end{equation}
where $\Tilde{\bm{K}} = \bm{K}\cdot\bm{R}$ and $\bm{K}$ is the kernel of the pupil model $\bm{A}$ as described in \citet{martinache2010}. We also decided to normalize the rows of the $\Tilde{\bm{K}}$ matrix to one so that the amplitude of the kernel phase signal is no longer depending on the geometry of the pupil model. This enables a more direct comparison between kernel phase observables obtained from different telescopes/instruments. To analytically estimate the kernel phase uncertainties, we first need to linearize the relationship between the kernel phase $\theta$ and the image $I$, so that
\begin{equation}
    \theta \approx \bm{B}\cdot I = \Tilde{\bm{K}}\cdot\left(\frac{\mathfrak{Im}(\bm{F})}{\left|\bm{F}\cdot I\right|}\right)\cdot I,
\end{equation}
where $\bm{F}$ denotes the linear discrete Fourier transform at the spatial frequencies of the pupil model and the fraction denotes element-wise division \citep{kammerer2019}. Then, the kernel phase covariance $\bm{\Sigma}$ can be computed from the image covariance $\bm{\Sigma}_\text{image}$ as
\begin{equation}
    \label{eqn:kpcov}
    \bm{\Sigma} = \bm{B}\cdot\bm{\Sigma}_\text{image}\cdot\bm{B}^T,
\end{equation}
where $\Sigma_{\text{image}}$ is assumed to be uncorrelated with the square of the ``ERR'' extension of the stage 2-reduced pipeline product on the diagonal.

\begin{figure*}[t!]
\centering
\includegraphics[width=\textwidth]{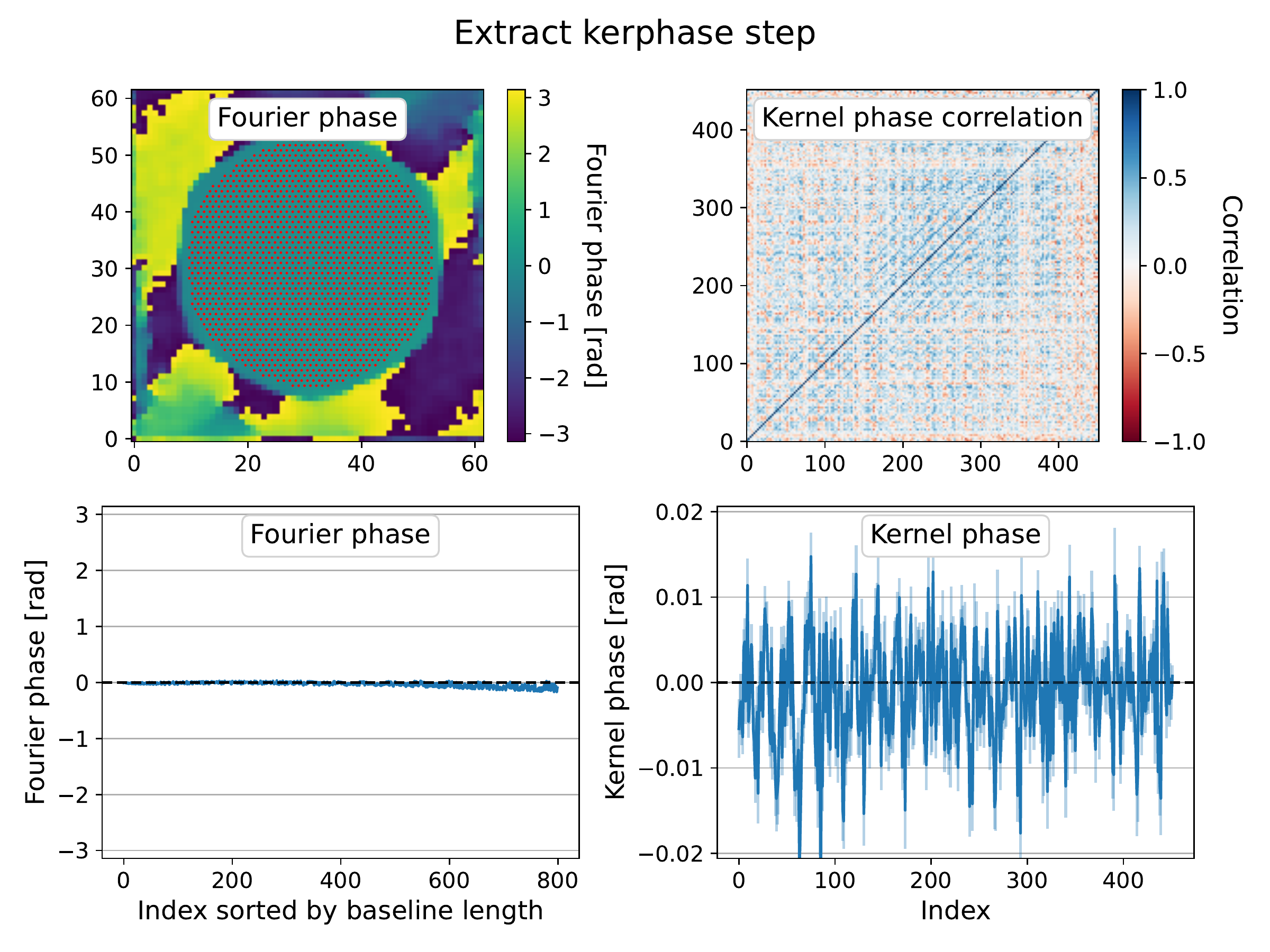}
\caption{Diagnostic plot produced by the kernel phase extraction step of the \texttt{Kpi3Pipeline} for the same data as shown in Figure~\ref{fig:bad_pixel_fixing}. The upper left panel shows the image Fourier phase together with the discrete spatial frequencies of the pupil model (red dots). The bottom panels show the image Fourier phase extracted at the discrete spatial frequencies of the pupil model (left) and the resulting kernel phase (right). In both panels, the theoretical signal of an unresolved point-source is constantly zero and indicated by a dashed black line. The upper right panel shows the kernel phase correlation matrix obtained from a linear propagation of the image pixel uncertainties (cf. Equation~\ref{eqn:kpcov}).}
\label{fig:kernel_phase_extraction}
\end{figure*}

If the recentering step is being run in conjunction with the kernel phase extraction step, a more accurate method is used to extract the kernel phase observables where the integer pixel recentering is performed by rolling the image and the subpixel recentering is performed by adding a linear ramp to the extracted image Fourier phase. This method circumvents interpolation errors when performing subpixel shifts of the images. We note that for extracting the kernel phase uncertainties, we still use the images that have been recentered with subpixel precision. The diagnostic plot of the kernel phase extraction step is shown in Figure~\ref{fig:kernel_phase_extraction}.

\subsubsection{Empirical uncertainties step}

A single \emph{JWST} data product (here called an exposure) typically consists of a large number ($\gtrsim10$) of individual integrations. While all previous pipeline steps can be performed on hundreds of individual images within timescales of seconds to minutes, it can be very time- and memory-consuming to perform model fitting on such a large number of individual data points, especially if accounting for correlated uncertainties. Hence, it is advisable to average individual integrations within a single exposure into one final data point. This is achieved in the \texttt{Kpi3Pipeline} by computing the covariance-weighted mean of the kernel phase observables according to
\begin{equation}
    \theta_\text{mean} = \bm{\Sigma}_\text{mean}\cdot\sum_{i=1}^{N_\text{f}}\bm{\Sigma}_i^{-1}\cdot\theta_i,
\end{equation}
where $\theta_i$ and $\bm{\Sigma}_i$ are the kernel phase and its covariance of the individual integrations and
\begin{equation}
    \bm{\Sigma}_\text{mean} = \left(\sum_{i=1}^{N_\text{f}}\bm{\Sigma}_i^{-1}\right)^{-1}
\end{equation}
is the mean covariance matrix \citep{kammerer2019}.

\begin{figure*}[t!]
\centering
\includegraphics[trim={0cm 23cm 0cm 0cm},clip,width=\textwidth]{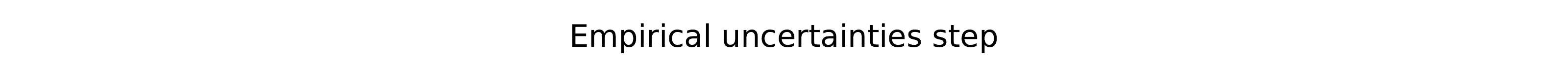}
\includegraphics[trim={0cm 0cm 0cm 12.5cm},clip,width=\textwidth]{figures/jw01093023001_03103_00002_nis_calints_emp_kpfits.pdf}
\caption{Diagnostic plot produced by the empirical uncertainties step of the \texttt{Kpi3Pipeline} for the same data as shown in Figure~\ref{fig:bad_pixel_fixing}. The minimum/maximum range over the exposure is shown as a light blue area and the weighted mean including uncertainties is shown as a solid blue line.}
\label{fig:empirical_uncertainties}
\end{figure*}

Another advantage of having several individual integrations is that the uncertainties can be estimated empirically as the standard deviation of the kernel phase observables over these individual integrations. This can be especially helpful if the kernel phase observables are dominated by systematic errors \citep[e.g.,][]{kammerer2019,wallace2020} which are not reflected in the analytical error estimation from the previous pipeline step. An empirical estimation for the kernel phase covariance can then be obtained by computing the correlation matrix $\bm{C}_\text{mean}$ of $\bm{\Sigma}_\text{mean}$ (which only includes image pixel uncertainties) and multiplying it with the empirically estimated kernel phase standard deviation $\sigma_\text{emp}$, that is
\begin{equation}
    \bm{\Sigma}_{\text{emp},nm} = \bm{C}_{\text{mean},nm}\sigma_{\text{emp},m}\sigma_{\text{emp},n}.
\end{equation}
Here, $m$ and $n$ denote the indices of the matrices. This procedure ensures that the kernel phase uncertainties remain consistent with the square root of the diagonal of the kernel phase covariance matrix. The diagnostic plot of the empirical uncertainties step is shown in Figure~\ref{fig:empirical_uncertainties}.

\subsection{Kernel phase FITS files}
\label{sec:kernel_phase_fits_files}

A common and well-defined data exchange format has proven highly valuable for sharing astronomical data and reproducing observational results. In the optical and near-infrared (long-baseline) interferometry community, the OIFITS file format \citep{pauls2005,duvert2017} has established itself as a golden standard that is used by most major observatories and instruments (e.g., CHARA/MIRC, \citealt{monnier2004}, VLTI/PIONIER, \citealt{lebouquin2011}, VLTI/GRAVITY, \citealt{lapeyrere2014}) and data analysis tools (e.g., \texttt{LITpro}, \citealt{tallon-bosc2008}, \texttt{CANDID}, \citealt{gallenne2015}). Due to the very similar nature of the data, the same OIFITS file format is also being used for AMI data (e.g., in \texttt{ImPlaneIA}, \citealt{greenbaum2015}, \texttt{AMICAL}, \citealt{soulain2020}, \texttt{SAMpip}\footnote{\url{https://github.com/cosmosz5/CASSINI}}). Although KPI and AMI are closely related techniques \citep[see, e.g.,][]{ireland2016}, there are a few subtle differences that strongly motivate the need for a distinct file format for KPI data. Firstly, as shown in Section~\ref{sec:introduction}, the kernel phase $\theta$ is obtained from a special linear combination $\Tilde{\bm{K}}$ of the image Fourier phase $\phi$. Thus, the kernel matrix $\Tilde{\bm{K}}$ is of fundamental importance in KPI and is repeatedly used in the model fitting process to project image Fourier phase onto kernel phase. Fast and easy access to this kernel matrix $\Tilde{\bm{K}}$ is therefore vital. While it is in principle possible (albeit time-consuming) to recompute this matrix from the geometry of the pupil array that can be saved in the OI\_ARRAY extension of an OIFITS file, we strongly prefer to save this matrix directly in one of the FITS file extensions, not least because handling hundreds of individual subapertures becomes impractical using the OI\_ARRAY extension. Secondly, recent work from \citet{martinache2020} has shown that the use of ``gray'' pupil models featuring subapertures with a continuous transmission $0 < T \leq 1$ is highly desirable in KPI as they help to reduce the systematic kernel phase signal of unresolved and point-like PSF reference targets. However, gray pupil models are not supported by the OIFITS file format due to the third dimension required to store transmission. Finally, the Karhunen-Lo\`eve calibration method applied in \citet{kammerer2019}, \citet{wallace2020}, and \citet{kammerer2021} can be used implicitly if the kernel phase $\theta$ and the kernel matrix $\Tilde{\bm{K}}$ are saved in matrix format. This is because the Karhunen-Lo\`eve calibration \citep{soummer2012} is a linear projection represented by a matrix $\bm{P}$ so that replacing the kernel phase and the kernel matrix with $\bm{P}\cdot\theta$ and $\bm{P}\cdot\Tilde{\bm{K}}$, respectively, will ensure that any model fitting code will automatically run correctly with a calibrated dataset.

Based on the aforementioned issues with the OIFITS file format, the kernel phase community has expressed the need for a dedicated data exchange format in the framework of the Masking/Kernel Phase Hackathon\footnote{\url{https://sites.google.com/uci.edu/virtualmaskinghackathon/home?authuser=1}} held virtually in mid 2021 to ensure compatibility among a variety of data reduction, calibration, and model fitting tools. The kernel phase FITS (KPFITS) file format proposed here is based on the \texttt{XARA} file format developed by \citet{martinache2010} and \citet{martinache2013}. However, based on recent developments from \citet{kammerer2019} and \citet{martinache2020} several modifications were made to the original \texttt{XARA} file format in consultation with kernel phase experts from the global community. For completeness, we also added an optional extension for saving kernel amplitude observables as introduced in \citet{pope2016b}. The structure of the KPFITS file format is described in Table~\ref{tab:kpfits} and shall be used as a standard for saving and exchanging kernel phase data in the future. The exact order of the FITS file extensions is in principle irrelevant as extensions shall be referred to in the code with their extension names, the numbering in Table~\ref{tab:kpfits} can hence be regarded a suggestion. We note that in principle, AMI data can also be exchanged using the KPFITS file format in a more efficient way.

To ensure that the information necessary to consistently reprocess the kernel phase data is present, the KPFITS file format also requires a PRIMARY FITS header with the header keywords specified in Table~\ref{tab:kpfits_header}. While several of these keywords are optional, others are strictly required to rerun the \texttt{Kpi3Pipeline} and extract kernel phase observables or to aid the model fitting procedures described in Section~\ref{sec:model_fitting_with_fouriever}.

\begin{table*}[t!]
\caption{Structure of the KPFITS file format. ``Ind'' denotes the index of the FITS file extension, ``Opt'' specifies whether an extension is optional or not, $N_\text{f}$ denotes the number of frames, $N_\lambda$ denotes the number of spectral channels, $N_\text{pix}$ denotes the number of pixels per image axis, $N_\text{sap}$ denotes the number of subapertures, $N_\text{uv}$ denotes the number of uv-points and $N_\text{ker}$ denotes the number of kernels. We note that the telescope images are optional and the PRIMARY extension can in principle be an empty array.}
\centering
\begin{tabular}{llllll}
Ind & Opt & Name & Type & Dimensions & Description\\
\hline
\hline
0 & (N) & PRIMARY & Image & $N_\text{f}\times N_\lambda\times N_\text{pix}\times N_\text{pix}$ & Telescope images (can be empty array)\\
\hline
1 & ~N & APERTURE & Table & $N_\text{sap}\times3$ & Description of pupil model\\
& ~N & ~~~XXC & Column & & Subaperture x-coordinate [m]\\
& ~N & ~~~YYC & Column & & Subaperture y-coordinate [m]\\
& ~N & ~~~TRM & Column & & Subaperture transmission ($0 < T \leq 1$)\\
\hline
2 & ~N & UV-PLANE & Table & $N_\text{uv}\times3$ & Fourier plane coverage of pupil model\\
& ~N & ~~~UUC & Column & & Fourier u-coordinate [m]\\
& ~N & ~~~VVC & Column & & Fourier v-coordinate [m]\\
& ~N & ~~~RED & Column & & Redundancy of uv-position (integer)\\
\hline
3 & ~N & KER-MAT & Image & $N_\text{ker}\times N_\text{uv}$ & \hspace{-1.4cm}\makecell[l]{Matrix $\Tilde{\bm{K}}$ mapping image Fourier phase\\onto kernel phase}\\
\hline
4 & ~N & BLM-MAT & Image & $N_\text{uv}\times N_\text{sap}$ & \hspace{-1.4cm}\makecell[l]{Matrix $\bm{A}$ mapping pupil plane phase\\onto image Fourier phase}\\
\hline
5 & ~N & KP-DATA & Image & $N_\text{f}\times N_\lambda\times N_\text{ker}$ & Kernel phase data [rad]\\
\hline
6 & ~N & KP-SIGM & Image & $N_\text{f}\times N_\lambda\times N_\text{ker}$ & Kernel phase uncertainties [rad]\\
\hline
7 & ~N & CWAVEL & Table & $N_\lambda\times2$ & Description of bandpass\\
& ~N & ~~~CWAVEL & Column & & Central wavelength of bandpass [m]\\
& ~N & ~~~BWIDTH & Column & & \hspace{-1.4cm}\makecell[l]{Best available estimate of effective half-\\power bandwidth [m]}\\
\hline
8 & ~N & DETPA & Image & $N_\text{f}$ & Detector position angle E of N [deg]\\
\hline
9 & ~N & CVIS-DATA & Image & $2\times N_\text{f}\times N_\lambda\times N_\text{uv}$ & \hspace{-1.4cm}\makecell[l]{Complex visibility data\\(dim 1 = real part, dim 2 = imag. part)}\\
\hline
$>9$ & ~Y & KA-DATA & Image & $N_\text{f}\times N_\lambda\times N_\text{ker}$ & Kernel amplitude data\\
\hline
$>9$ & ~Y & KA-SIGM & Image & $N_\text{f}\times N_\lambda\times N_\text{ker}$ & Kernel amplitude uncertainties\\
\hline
$>9$ & ~Y & CAL-MAT & Image & $(N_\text{ker}-K_\text{klip})\times N_\text{ker}$ & \hspace{-1.4cm}\makecell[l]{Karhunen-Lo\`eve projection matrix $\bm{P}'$\\as in K19}\\
\hline
$>9$ & ~Y & KP-COV & Image & \hspace{-1.4cm}\makecell[l]{Flexible, up to\\$N_\text{f}\times N_\lambda\times N_\text{ker}\times N_\text{ker}$} & Kernel phase covariance [rad${}^2$]\\
\hline
$>9$ & ~Y & KA-COV & Image & \hspace{-1.4cm}\makecell[l]{Flexible, up to\\$N_\text{f}\times N_\lambda\times N_\text{ker}\times N_\text{ker}$} & Kernel amplitude covariance\\
\hline
$>9$ & ~Y & FULL-COV & Image & \hspace{-1.4cm}\makecell[l]{Flexible, up to\\$N_\text{f}\times N_\lambda\times2N_\text{ker}\times2N_\text{ker}$} & \hspace{-1.4cm}\makecell[l]{Kernel phase [rad${}^2$] and kernel amplitude\\covariance}\\
\hline
$>9$ & ~Y & IMSHIFT & Table & $N_\text{f}\times2$ & Shift to recenter images\\
& ~Y & ~~~XSHIFT & Column & & Shift along x-axis (1-axis in Python) [pix]\\
& ~Y & ~~~YSHIFT & Column & & Shift along y-axis (0-axis in Python) [pix]\\
\hline
$>9$ & ~Y & WINMASK & Image & $N_\text{pix}\times N_\text{pix}$ & Super-Gaussian windowing mask\\
\hline
\multicolumn{6}{l}{\textbf{Notes.} K19 = \citet{kammerer2019}. All data should be of type float.}
\end{tabular}
\label{tab:kpfits}
\end{table*}

\begin{table*}[t!]
\caption{PRIMARY header of the KPFITS file format. ``Opt'' specifies whether a header keyword is optional or not.}
\centering
\begin{tabular}{lllll}
Opt & Name & Type & Unit or value & Description\\
\hline
\hline
N & PSCALE & float & mas/pix & Detector pixel scale\\
N & GAIN & float & ADU/e- & Detector gain\\
N & DIAM & float & m & Primary mirror diameter\\
N & EXPTIME & float & s & Exposure time per frame\\
N & TELESCOP & str & e.g., JWST & Telescope name\\
N & INSTRUME & str & e.g., NIRISS & Instrument name\\
Y & DATEOBS & str & YYYY-MM-DDTHH:MM:SS & Date of observation\\
Y & TARGNAME & str & e.g., AX Cir & Target name\\
Y & TARG\_RA & float & deg & Target RA (mid exposure)\\
Y & TARG\_DEC & float & deg & Target DEC (mid exposure)\\
Y & FILTER & str & e.g., F480M & Filter name\\
Y & POLCHANN & str & -- & Polarization channel\\
Y & PATTTYPE & str & e.g., NONE/5-POINT-BOX & Dither pattern name\\
Y & PATT\_NUM & int & e.g., 1 & Position number in dither pattern\\
Y & NUMDTHPT & int & e.g., 5 & Total number of positions in dither pattern\\
Y & PROCSOFT & str & e.g., XARA & Processing software name\\
Y & WRAD & float & pix & Radius of super-Gaussian windowing mask\\
Y & CALFLAG & str & True/False & Has the data been calibrated?\\
N & CONTENT & str & KPFITS1 & Name of file format\\
\hline
\end{tabular}
\label{tab:kpfits_header}
\end{table*}

\subsection{Model fitting with \texttt{fouriever}}
\label{sec:model_fitting_with_fouriever}

There are several publicly available model fitting toolkits that understand OIFITS files and can be used to analyze long-baseline interferometry and AMI data (e.g., \texttt{LITpro}, \citealt{tallon-bosc2008}, \texttt{CANDID}, \citealt{gallenne2015}, \texttt{PMOIRED}, \citealt{merand2022}). However, none of these toolkits accounts for correlated uncertainties in the fits. Recently, \citet{lachaume2019} and \citet{kammerer2020} developed methods to extract and model correlations in long-baseline interferometry data and \citet{kammerer2020} showed that accounting for them in the fits can improve VLTI/GRAVITY companion detection limits by a factor of up to $\sim2$. More importantly, they also found that widely used detection criteria based on $\chi^2$-statistics are only valid when accounting for correlations and yield an excessive number of false positive detections otherwise.

In light of these findings, we develop the \texttt{fouriever} toolkit which provides a single solution for analyzing KPI, AMI, and long-baseline interferometry data while accounting for correlated uncertainties in the fits. The toolkit also enables modeling correlations in long-baseline interferometry and AMI data as described in \citet{kammerer2020} and calibrating science data using a Karhunen-Lo\`eve projection based on calibrator data as introduced in \citet{kammerer2019}. The \texttt{fouriever} toolkit is written in Python and can be obtained from GitHub\footnote{\url{https://github.com/kammerje/fouriever}}. Many functionalities were inspired by \texttt{CANDID} but have been modified to improve the performance given that fits accounting for correlations involve significantly more complex matrix multiplications. In the following, we briefly outline the search for companions and the estimation of detection limits with \texttt{fouriever} and highlight similarities and improvements with respect to \texttt{CANDID}. We also note that \texttt{fouriever} comes with tutorials and test data for \emph{JWST} NIRISS AMI, VLT NACO and Keck NIRC2 KPI, and VLTI PIONIER and VLTI GRAVITY long-baseline interferometry applications.

\subsubsection{Companion search}
\label{sec:companion_search}

The first step in the analysis of high-contrast imaging data usually consists of the search for one or multiple companions. For this purpose, \texttt{fouriever} can compute a $\chi^2$-detection map based on binary model fits similar to the \texttt{fitMap} in \texttt{CANDID} but accounting for correlated uncertainties in the fits. This $\chi^2$-detection map is obtained from a set of least squares gradient descent minimizations initialized on a grid around the science target. The optimizer aims to minimize
\begin{equation}
    \chi^2 = R^T\cdot\bm{\Sigma}^{-1}\cdot R,
\end{equation}
where $R = D-M$ are the residuals between data and model and $\bm{\Sigma}$ is the data covariance matrix. For simplicity, other toolkits assume that there are no correlations and that the matrix $\bm{\Sigma}$ is diagonal. The model observables are obtained from the complex visibility of a binary source
\begin{equation}
    V_\text{bin} = \frac{V_1+V_2f\exp\left(-2\pi i\left(\frac{\Delta_\text{RA}u}{\lambda}+\frac{\Delta_\text{DEC}v}{\lambda}\right)\right)}{1+f},
\end{equation}
where $V_1$ and $V_2$ are the complex visibilities of the primary and the secondary, $f$ is the relative flux of the secondary with respect to the primary, $\Delta_\text{RA}$ and $\Delta_\text{DEC}$ are the right ascension and declination offset between the primary and the secondary, $u$ and $v$ are the Fourier plane coordinates for which the complex visibility shall be obtained, and $\lambda$ is the observing wavelength \citep[e.g.,][]{berger2003}. For the case of an unresolved companion considered here, we set
\begin{align}
    V_1 &= 2\frac{J_1(\pi\vartheta b)}{\pi\vartheta b},\\
    V_2 &= 1,
\end{align}
where $J_1$ denotes the first-order Bessel function of the first kind and $b = \sqrt{u^2+v^2}$ denotes the length of the baseline for which the complex visibility shall be obtained, so that the primary is described by a uniform disk with angular diameter $\vartheta$ and the secondary is described by an unresolved point-source \citep[e.g.,][]{berger2003}. The model observables $M$ in the form of squared visibility amplitudes (v2), closure phases (cp), or kernel phases (kp) are then obtained via
\begin{align}
    \text{v2} &= \left|V\right|^2,\\
    \text{cp} &= \bm{C}\cdot\angle V,\\
    \text{kp} &= \Tilde{\bm{K}}\cdot\angle V,
\end{align}
where $\angle$ denotes the phase of a complex number and $\bm{C}$ and $\Tilde{\bm{K}}$ denote the closure and kernel matrices, respectively, whose rows contain the linear combinations of Fourier phases forming closure and kernel phases. The detection significance is obtained using $\chi^2$-statistics and computed via
\begin{equation}
    P_\text{bin} = 1-\text{CDF}_\nu\left(\frac{\nu\chi^2_\text{r,ud}}{\chi^2_\text{r,bin}}\right),
    \label{eqn:p_bin}
\end{equation}
where $\text{CDF}_\nu$ denotes the cumulative distribution function of a $\chi^2$-distribution with $\nu$ degrees of freedom, $\chi^2_\text{r,ud}$ is the reduced $\chi^2$ of the uniform disk only model (i.e., without a companion), and $\chi^2_\text{r,bin}$ is the reduced $\chi^2$ of the binary model. As in \texttt{CANDID}, \texttt{fouriever} also supports numerical bandwidth smearing which helps to prevent underestimating the companion flux. Moreover, an on-sky search region can be specified where \texttt{fouriever} is looking for companions. This feature is particularly useful if an already known companion resides outside the diffraction field-of-view (FoV) of $0.5\lambda_\text{min}/B_\text{min}$ and creates aliasing artifacts that could be mistaken for a true companion, where $\lambda_\text{min}$ is the shortest observing wavelength and $B_\text{min}$ is the smallest baseline of the pupil or interferometric array. Searching for additional companions is possible by repeating the computation of the $\chi^2$-detection map after analytically subtracting the best fit companion model from the data.

The uncertainties derived from least squares gradient descent minimizations are usually unreliable if the uncertainties in the underlying data have been wrongly estimated. While \texttt{CANDID} employs the bootstrapping (with replacement) method \citep{efron1986} to extract the uncertainties and correlations of the model parameters, \texttt{fouriever} employs the MCMC method \citep[using \texttt{emcee},][]{foreman-mackey2013} with a temperature $T_{\chi^2}$ by default equaling the reduced $\chi^2$ of the best fit binary model obtained from the $\chi^2$-detection map to account for potential over- or underestimation of the uncertainties in the underlying data \citep[e.g.,][]{andrae2010}. The log-likelihood function that is being sampled by the MCMC method is thus given by
\begin{equation}
    \log\mathcal{L} = -\frac{1}{2}\frac{\chi^2}{T_{\chi^2}}.
\end{equation}
Both the bootstrapping and the MCMC method yield comparable results and an advantage of the latter is that it can also be used to fit more complex models such as multiple companions simultaneously or geometric disk and ring models to the data \citep[see, e.g.,][]{kammerer2021,blakely2022}. 

Figure~\ref{fig:axcir_companion} shows a companion search in VLTI/PIONIER data of AX~Cir from \citet{gallenne2015} with \texttt{fouriever} and \texttt{CANDID}. For a more direct comparison, we did assume uncorrelated uncertainties and numerical bandwidth smearing evaluated at three uniformly spaced wavelength nodes across the observing bandpass in both cases. The recovered model parameters (host star uniform disk diameter $\vartheta$/diam*, relative companion flux $f$, right ascension offset of the companion $\Delta_\text{RA}$/$x$, and declination offset of the companion $\Delta_\text{DEC}$/$y$) from the MCMC fit with \texttt{fouriever} and the bootstrapping with \texttt{CANDID} agree well within their uncertainties. Both approaches also reveal a correlation in the model parameters between the host star uniform disk diameter and the relative companion flux. We observe a similar computation time for \texttt{fouriever} and \texttt{CANDID}, although we note that \texttt{fouriever} natively uses a higher resolution grid than \texttt{CANDID} when computing the $\chi^2$-detection map and also achieves a similar performance with correlated uncertainties.

\subsubsection{Detection limits}
\label{sec:detection_limits}

Estimating companion detection limits (also known as contrast curves) is one of the most fundamental pathways to assess the sensitivity of high-contrast imaging observations and to quantitatively interpret non-detections. For this purpose, \texttt{fouriever} offers the same two methods to estimate these limits as \texttt{CANDID}, namely the ``Absil'' and the ``Injection'' method.

\noindent\textbf{``Absil'' method:} this method has been introduced by \citet{absil2011} and refined by \citet{gallenne2015} and computes the detection limit as the relative companion flux $f$ at which the binary model deviates by a certain probability (e.g., 99.73\% for 3--$\sigma$) from the uniform disk only model according to
\begin{equation}
    P_\text{det} = 1-\text{CDF}_\nu\left(\frac{\nu\chi^2_\text{r,bin}}{\chi^2_\text{r,ud}}\right),
\end{equation}
that is assuming that the binary model is the true model (see \citealt{gallenne2015} for a more detailed discussion of this choice). This condition is evaluated on a grid around the science target and azimuthally averaged to obtain a contrast curve.

\noindent\textbf{``Injection'' method:} this method has been introduced by \citet{gallenne2015} and has been reported to be more robust than the ``Absil'' method if the data is biased or affected by correlations. It consists of analytically injecting a fake companion into the data and then computing the significance of the best fit uniform disk only model over the best fit binary model fitted to the synthetic data with the injected companion. In this case, $P_\text{bin}$ (Equation~\ref{eqn:p_bin}) sets the detection limit in the form of relative companion flux $f$. This method is computationally more expensive than the ``Absil'' method but it resembles more closely the procedure that is undertaken if a real companion is detected.

\begin{table*}[t!]
\caption{\emph{JWST} NIRISS full pupil KPI observations taken as part of the instrument commissioning on 23 May 2022 and 5 June 2022 (PID~1093, PI Deepashri Thatte). BP denotes the fractional filter bandpass, $N_\text{exp}$ is the number of exposures, $N_\text{int}$ is the number of individual integrations per exposure, $N_\text{group}$ is the number of groups per integration, $T_\text{int}$ is the effective integration time, and $T_\text{exp}$ is the effective exposure time (per individual exposure). Reobs = reobserved on 5 June 2022.}
\centering
\begin{tabular}{llllllllll}
No. & Target name & Filter & BP & $N_\text{exp}$ & $N_\text{int}$ & $N_\text{group}$ & $T_\text{int}$ [s] & $T_\text{exp}$ [s] & Reobs \\
\hline
\hline
1 & 2MASS~J062802.01-663738.0 & F480M & 6.2\% & 2 & 239 & 14 & 1.05616 & 252.288 & Y\\
2 & TYC~8906-1660-1 & F480M & 6.2\% & 2 & 236 & 13 & 0.98072 & 231.327 & Y\\
3 & CPD-67~607 & F480M & 6.2\% & 2 & 232 & 13 & 0.98072 & 227.406 & Y\\
4 & CPD-66~562 & F480M & 6.2\% & 2 & 241 & 14 & 1.05616 & 254.400 & N\\
\hline
\end{tabular}
\label{tab:observations}
\end{table*}

The \texttt{fouriever} toolkit can compute these detection limits accounting for correlated uncertainties in the fits and uses a more aggressive smoothing when azimuthally averaging the detection maps because the sparse uv-coverage especially in long-baseline interferometry observations often results in detection maps showing strong changes in sensitivity at high spatial frequencies (aliasing). This can be seen in the top panels of Figure~\ref{fig:axcir_detlims} which also compares the detection limits of the VLTI/PIONIER data of AX~Cir from \citet{gallenne2015} obtained with \texttt{fouriever} and \texttt{CANDID}. For better comparability, we apply the same azimuthal averaging that is being used in \texttt{fouriever} when computing the \texttt{CANDID} detection limits shown in Figure~\ref{fig:axcir_detlims}. To illustrate how accounting for correlated uncertainties in the fits improves the detection limits, we also model the correlations among the closure phase observables of the AX~Cir data considering that two telescope triplets of closing triangles at the VLTI always share one of their three baselines and are therefore not mathematically independent \citep[e.g.,][]{monnier2007}. This results in the correlation models from Section~2.2 of \citet{kammerer2020} with $x = y = 0$. Then, we repeat the computation of the detection limits with \texttt{fouriever} using these correlation models in the fits. For the AX~Cir data with only four closing triangles and three spectral channels, the correlations are small and the detection limits improve by only $\sim5\%$ (Figure~\ref{fig:axcir_detlims}).

\section{\emph{JWST} NIRISS observations}
\label{sec:observations}

At the high Strehl and the unparalleled thermal stability that can be achieved from space, \emph{JWST} is an ideal observatory for kernel phase imaging \citep{sivaramakrishnan2022}. Based on simulations, \citet{sallum2019} predict contrast limits of up to $\sim8$~mag in 90~min of observations (including overheads) at separations of $\gtrsim200$~mas with NIRCam which is comparable to the performance achieved by NIRISS AMI for bright targets and more than 1~mag better for faint targets. We note that for NIRISS KPI, we expect a similar performance as for NIRCam KPI \citep{ceau2019}.

Here, we analyze NIRISS CLEARP (full pupil) images that have been taken as part of the instrument commissioning on 23 May 2022 and 5 June 2022 (PID~1093, PI Deepashri Thatte). Besides the NRM, NIRISS is also equipped with a full pupil mask (Figure~\ref{fig:pupil_model}) that can be used for full pupil kernel phase or reference star direct imaging. Four targets were observed, each with two exposures consisting of 232--241 individual integrations resulting in an effective exposure time of $\sim227$--254~s (see Table~\ref{tab:observations}). Each exposure was designed to collect $1\mathrm{e}{8}$ photons on the detector according to the \emph{JWST} Exposure Time Calculator\footnote{\url{https://jwst.etc.stsci.edu/}}. Since the target acquisition did not work as expected during the first set of observations on 23 May 2022, three of the four targets were reobserved on 5 June 2022. The fourth target was not reobserved because a low-contrast close-in companion candidate was detected around it in the first set of observations so that the target is not useful for estimating companion detection limits for the NIRISS KPI mode. All data were processed with the \texttt{jwst} stage 1 and 2 pipelines and then fed into our custom kernel phase stage 3 pipeline as described in Section~\ref{sec:kernel_phase_stage_3_pipeline}. For the analysis presented here, we assume a central wavelength of $4.813019~\text{\textmu m}$\footnote{\url{http://svo2.cab.inta-csic.es/theory/fps/}} \citep{rodrigo2020} for the F480M bandpass and a detector pixel scale of 65.6~max/pix. Although we only analyze NIRISS data here, we note that the \texttt{Kpi3Pipeline} also supports NIRCam data and a pupil model for the NIRCam full pupil is provided.

\begin{figure}[t!]
\centering
\includegraphics[width=\columnwidth]{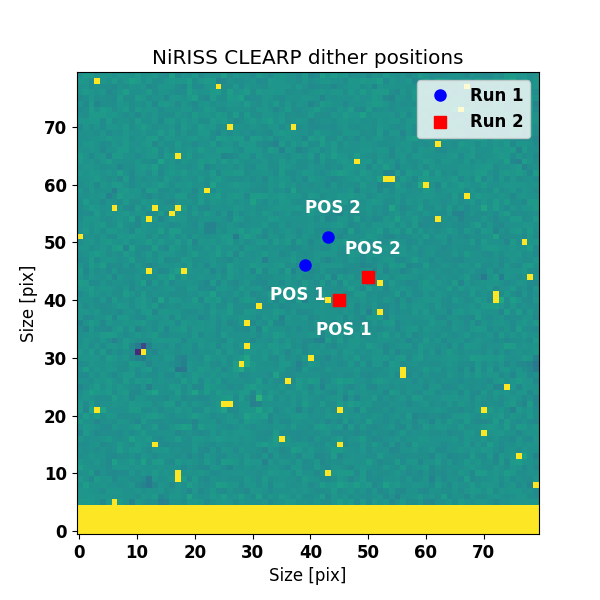}
\caption{NIRISS CLEARP full pupil imaging subarray with the two available dither positions POS~1 and POS~2 highlighted in red. Shown in blue are the two dither positions where the PSF was accidentally placed during the first of the two runs due to malfunctioning target acquisition. The pixels masked in yellow are static bad pixels from a flat image. The bottom five rows are reference pixels for correcting bias drifts which are not used for science.}
\label{fig:dither_positions}
\end{figure}

For the instrument commissioning, two different dither positions on the detector were explored: POS~1 and POS~2 (Figure~\ref{fig:dither_positions}). For each observed target, the first exposure uses POS~1 and the second exposure uses POS~2. However, due to malfunctioning target acquisition, the PSF was accidentally placed on different detector positions during the first of the two runs, so that in total four different dither positions were explored during instrument commissioning. The performance of these four different dither positions is analyzed in Section~\ref{sec:kpi_detection_limits}. We note that due to the high temporal stability of the observatory and the precise target acquisition procedure, dithering is not recommended for NIRISS KPI (and AMI) observations. Instead, it is recommended to select one dither position and use it for all science and reference star observations throughout the entire observing run.

\section{Results \& discussion}
\label{sec:results_and_discussion}

As mentioned in Section~\ref{sec:observations}, we report the discovery of a low-contrast close-in companion candidate around CPD-66~562 in Section~\ref{sec:a_low-contrast_close-in_companion_candidate_around_CPD-66_562}, a target that was chosen for the NIRISS instrument commissioning program because it was believed to be an isolated point-source. Moreover, we investigate the stability of the kernel phase observables throughout the observations and report faint source detection limits for the other three targets in Section~\ref{sec:kpi_detection_limits}.

\subsection{A low-contrast close-in companion candidate around CPD-66~562}
\label{sec:a_low-contrast_close-in_companion_candidate_around_CPD-66_562}

\begin{figure}[t!]
\centering
\includegraphics[width=\columnwidth]{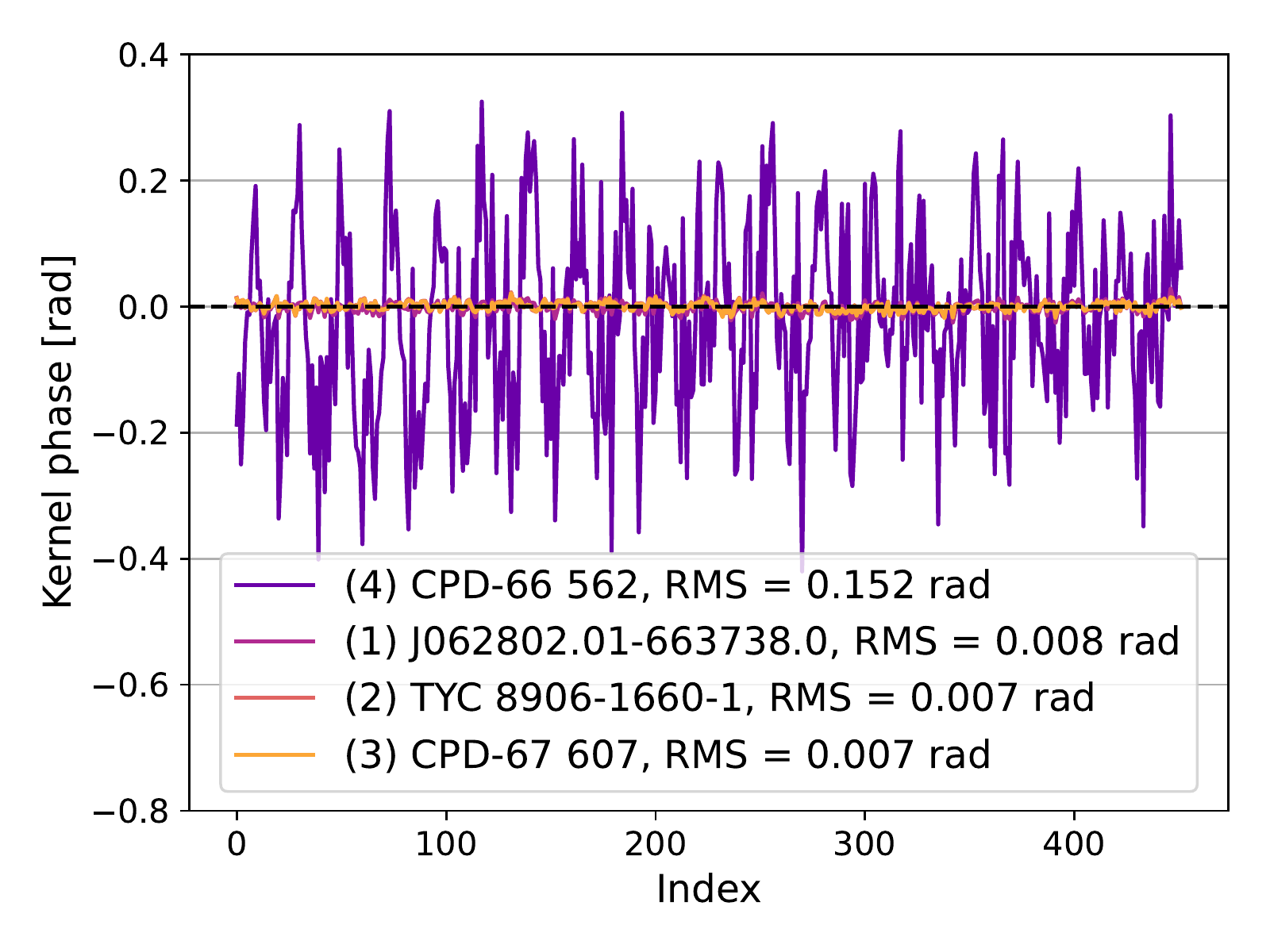}
\caption{Raw kernel phase signal of CPD-66~562 compared to that of the other three targets observed during \emph{JWST} NIRISS instrument commissioning. The kernel phase signal of CPD-66~562 shows a significantly larger scatter than that of the other three targets, indicating the detection of non-centro-symmetric emission such as from a low-contrast companion.}
\label{fig:raw_kerphase_run1}
\end{figure}

When inspecting the \texttt{jwst} pipeline-processed commissioning data by eye it is already possible to identify a low-contrast close-in companion candidate around CPD-66~562. This observation is supported by the raw kernel phase observables extracted with the \texttt{Kpi3Pipeline} which show a much larger scatter ($\pm0.152$~rad) for CPD-66~562 than for the other three targets ($\pm0.007$~rad, Figure~\ref{fig:raw_kerphase_run1}). We note that the theoretically expected kernel phase signal of a centro-symmetric point-source is zero so that larger scatter in the kernel phase observables is always indicative of a detection \citep[e.g.,][]{martinache2010}.

\begin{figure*}[t!]
\centering
\includegraphics[width=0.49\textwidth]{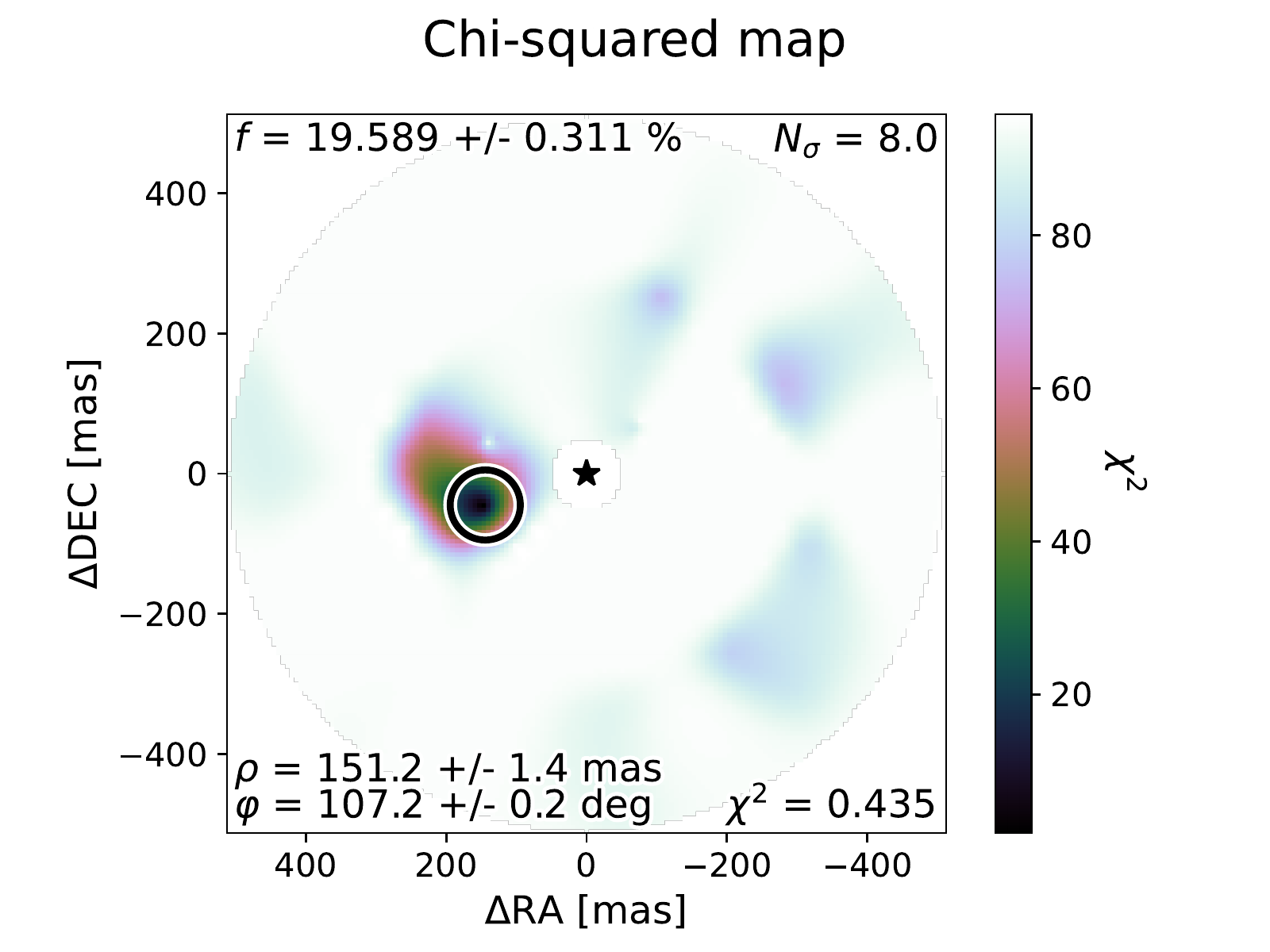}
\includegraphics[width=0.49\textwidth]{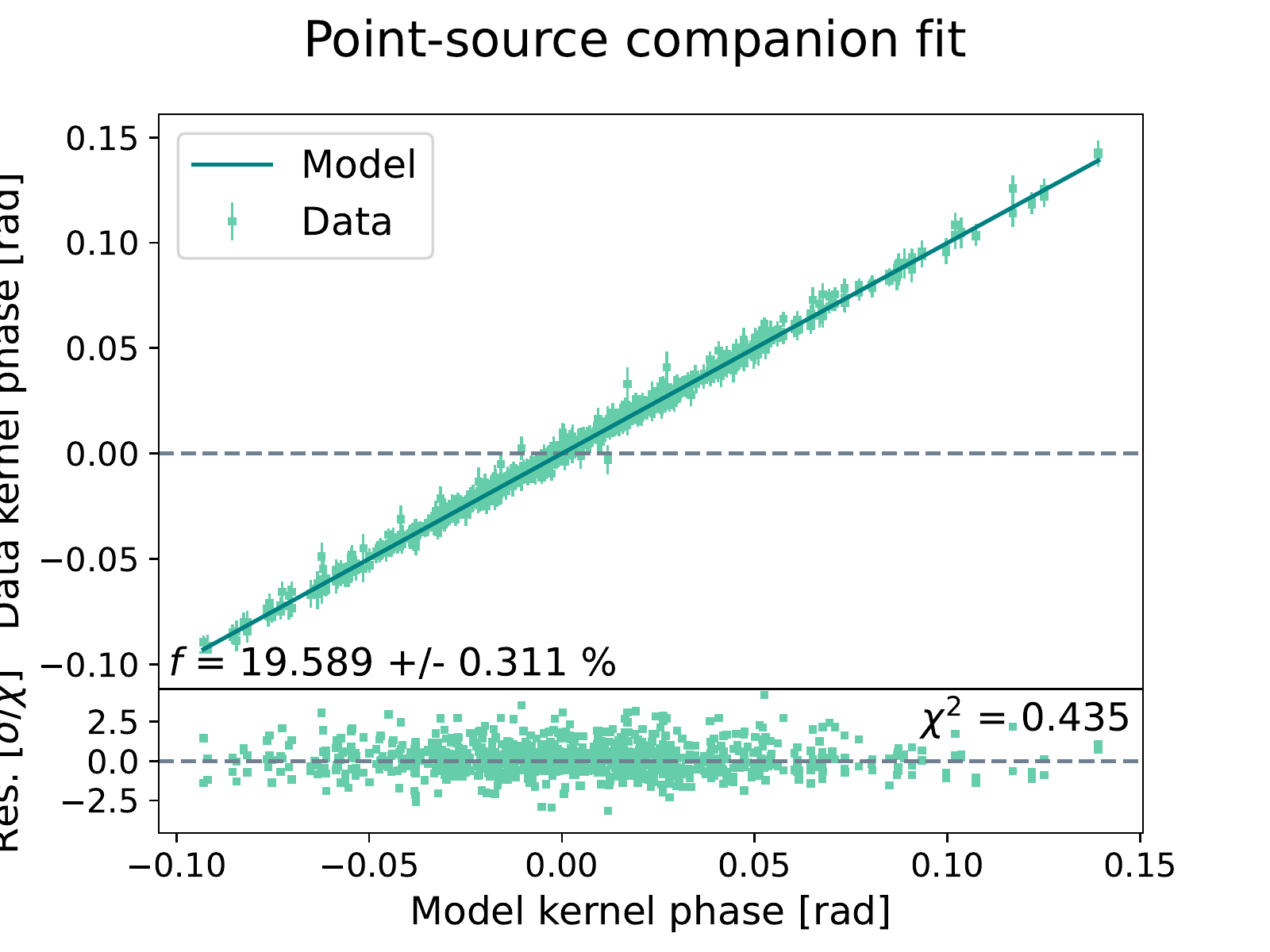}
\includegraphics[width=0.49\textwidth]{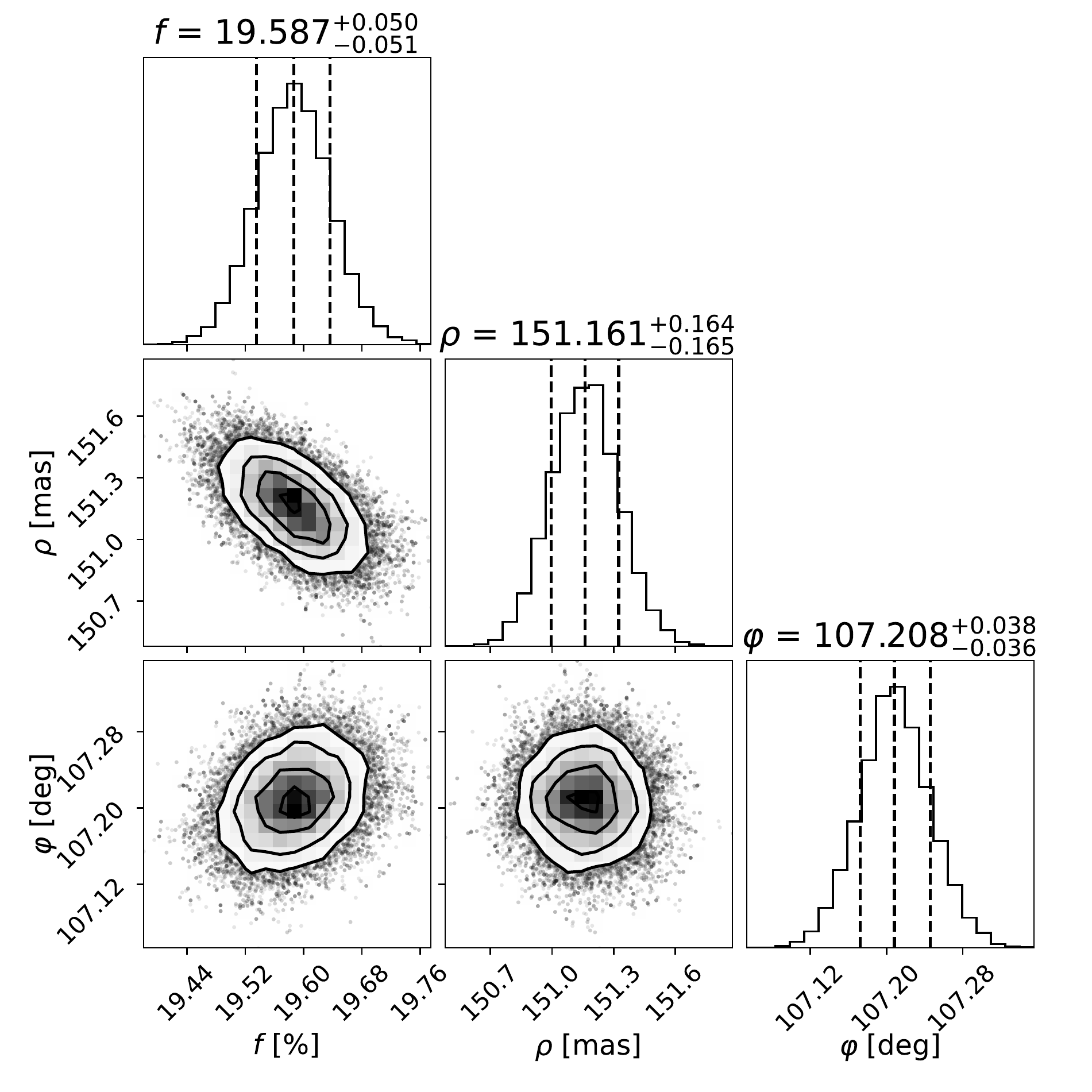}
\caption{Kernel phase detection of a low-contrast close-in companion candidate around CPD-66~562 in the \emph{JWST} NIRISS full pupil KPI commissioning data. The top two panels show the $\chi^2$-detection map and the data vs. model kernel phase signal of the best fit binary model obtained with \texttt{fouriever}. In the top left panel, the host star is located in the center and highlighted by a black star and the best fit companion position is highlighted by a black circle. The bottom panel shows the corner plot from an MCMC fit initialized around the best fit position from the $\chi^2$-detection map with more credible parameter uncertainties than the ones from the least-squares gradient descent minimization shown in the top panels.}
\label{fig:cpd-66_562_fouriever}
\end{figure*}

A companion search with \texttt{fouriever} reveals the precise parameters of the detected companion candidate. Before, though, we calibrate the raw kernel phase observables of CPD-66~562 using the other three targets as point-source references. Here, we consider the data from both dither positions of the first run on 23 May 2022. Using the Karhunen-Lo\`eve calibration class (\texttt{klcal}) in \texttt{fouriever}, the Karhunen-Lo\`eve basis of the kernel phase observables of the three reference targets is computed and clipped to $K_\text{klip} = 3$ principal components \citep{soummer2012,kammerer2019}. Then, the kernel phase observables of CPD-66~562 are projected into an $N_\text{ker}-3$-dimensional subspace that is orthogonal to (and thus independent of) these first three principal components of the reference target kernel phase observables. Then, the location of the companion candidate is derived by computing a $\chi^2$-detection map from the calibrated kernel phase observables of CPD-66~562 as described in Section~\ref{sec:companion_search}. Finally, the companion parameters including their uncertainties are estimated using an MCMC fit initialized around the best fit position from the $\chi^2$-detection map (Figure~\ref{fig:cpd-66_562_fouriever}).

The $\chi^2$-detection map shows a highly significant detection reaching the numerically set limit of $N_\sigma = 8.0$. The companion candidate is fairly bright with a relative flux of $\sim20\%$ of the primary and separated by $\sim151$~mas which corresponds to only $\sim1~\lambda/D$ at $\lambda = 4.8~\text{\textmu m}$. The small relative uncertainties in the flux of $<1\%$ and in the position of $<1$~permille obtained from the MCMC fit demonstrate the high precision at which the kernel phase technique can resolve companions at the diffraction limit.

\subsection{KPI detection limits}
\label{sec:kpi_detection_limits}

To evaluate the performance of \emph{JWST} NIRISS full pupil KPI, we compute 5--$\sigma$ companion detection limits using the Absil method in \texttt{fouriever}. For this purpose, we only consider the three point-source reference targets that were reobserved on 5 Jun 2022 with successful target acquisition achieving a precision of $<0.1$~pixels \citep{rigby2022}. This also means that we exclude the low-contrast binary CPD-66~562 reported in Section~\ref{sec:a_low-contrast_close-in_companion_candidate_around_CPD-66_562}.

\begin{figure*}[t!]
\centering
\includegraphics[width=\textwidth]{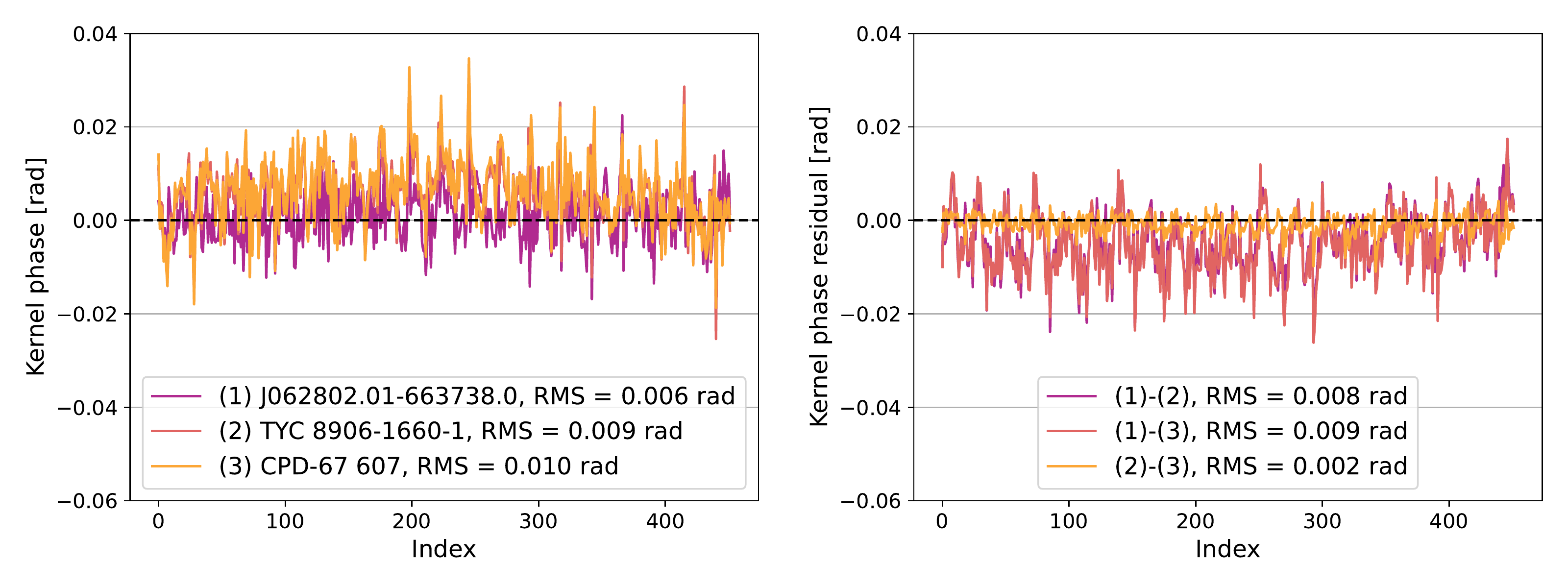}
\caption{Left: raw kernel phase signal of the three point-source reference targets observed during the second KPI run of \emph{JWST} NIRISS instrument commissioning. Right: kernel phase residuals between each pair of targets. Targets (2) and (3) show a larger systematic kernel phase signal than target (1), but their signals do also agree much better resulting in smaller kernel phase residuals between them and ultimately a much better calibration.}
\label{fig:raw_kerphase_run2}
\end{figure*}

First, we investigate the raw kernel phase signal of the three point-source reference targets from the second run averaged over both dither positions (Figure~\ref{fig:raw_kerphase_run2}). The average observed scatter is $\pm0.008$~rad and thus comparable, albeit slightly larger, to the first run. This small difference can stem from unflagged (and thus uncorrected) bad pixels in the data from the second run (e.g., from cosmic rays) or the temporally varying wavefront quality of \emph{JWST} \citep[due to e.g., wing tilt events,][]{rigby2022}. Since kernel phase observables are often dominated by systematic errors from higher order phase aberrations in the system \citep[e.g.,][]{kammerer2019,martinache2020}, it is common practice to use reference targets for calibrating out these systematic errors. The stability of the system can then be assessed by considering the difference between kernel phase observables from different reference targets. For this purpose, the right panel of Figure~\ref{fig:raw_kerphase_run2} shows the kernel phase residuals between each pair of targets observed in the second run (again averaged over both dither positions) and reveals that the signal measured on 2MASS~J062802.01-663738.0 seems to be an outlier.

Since two slightly different dither positions were used for the two exposures on each target, we repeat the same analysis for each dither position separately. Figure~\ref{fig:raw_kerphase_run2_dither} shows 2MASS~J062802.01-663738.0 as an outlier in both dither positions. A companion search with \texttt{fouriever} using the other two targets as calibrators reveals that this outlier is roughly consistent with a source at $\sim240$~mas separation and $\sim0.6\%$ flux ratio (Figure~\ref{fig:outlier_companion_search}). At a contrast of $\sim5.6$~mag, this companion candidate is above the 5--$\sigma$ detection threshold measured for the other two targets. To exclude the possibility that this detection is caused by wavefront drift between the observations of the different targets, we also analyze the data from the first run for an additional companion candidate around 2MASS~J062802.01-663738.0. Figure~\ref{fig:raw_kerphase_run1_dither} compares the kernel phase residuals between each pair of targets from the first run and also reveals 2MASS~J062802.01-663738.0 as an outlier in both dither positions. A companion search using only the data from the first run also yields a detection towards the South-East of 2MASS~J062802.01-663738.0, albeit at significantly different separation and flux ratio.

As a final check, we use classical PSF subtraction methods to search for the companion candidate in the image plane using only the data from the second run. We build a PSF library with the images of the other two targets (TYC~8906-1660-1 and CPD-67~607) and use the publicly available \texttt{pyKLIP} package \citep{wang2015} to subtract the first 20 Karhunen-Lo\`eve modes of the PSF library from the science target (2MASS~J062802.01-663738.0). The bottom panel of Figure~\ref{fig:outlier_companion_search} shows the PSF-subtracted image and clearly reveals a point-source at a position that is consistent with the best fit from the kernel phase reduction. Hence, the detection around 2MASS~J062802.01-663738.0 appears to be real and could be either a companion or a background source. The agreement between the kernel phase and the classical PSF subtraction techniques is a great confirmation of our methodology and a more detailed comparison between Fourier plane and classical imaging techniques is planned for a future publication.

Finally, comparing the four dither positions tried during instrument commissioning suggests that the first dither position from the second run should be avoided since it yields a $\sim3$~times larger scatter in the raw kernel phase and a $\sim2$~times larger scatter in the kernel phase residuals. We note that the RMS of the kernel phase uncertainty estimated from the standard deviation over the data cubes is on the order of $\sim0.0034$~rad for all targets and thus slightly larger than the calibration residuals between the best two targets, meaning that with $2\mathrm{e}{8}$ collected photons the observations are not yet limited by systematic calibration errors.

\begin{figure}[t!]
\centering
\includegraphics[width=\columnwidth]{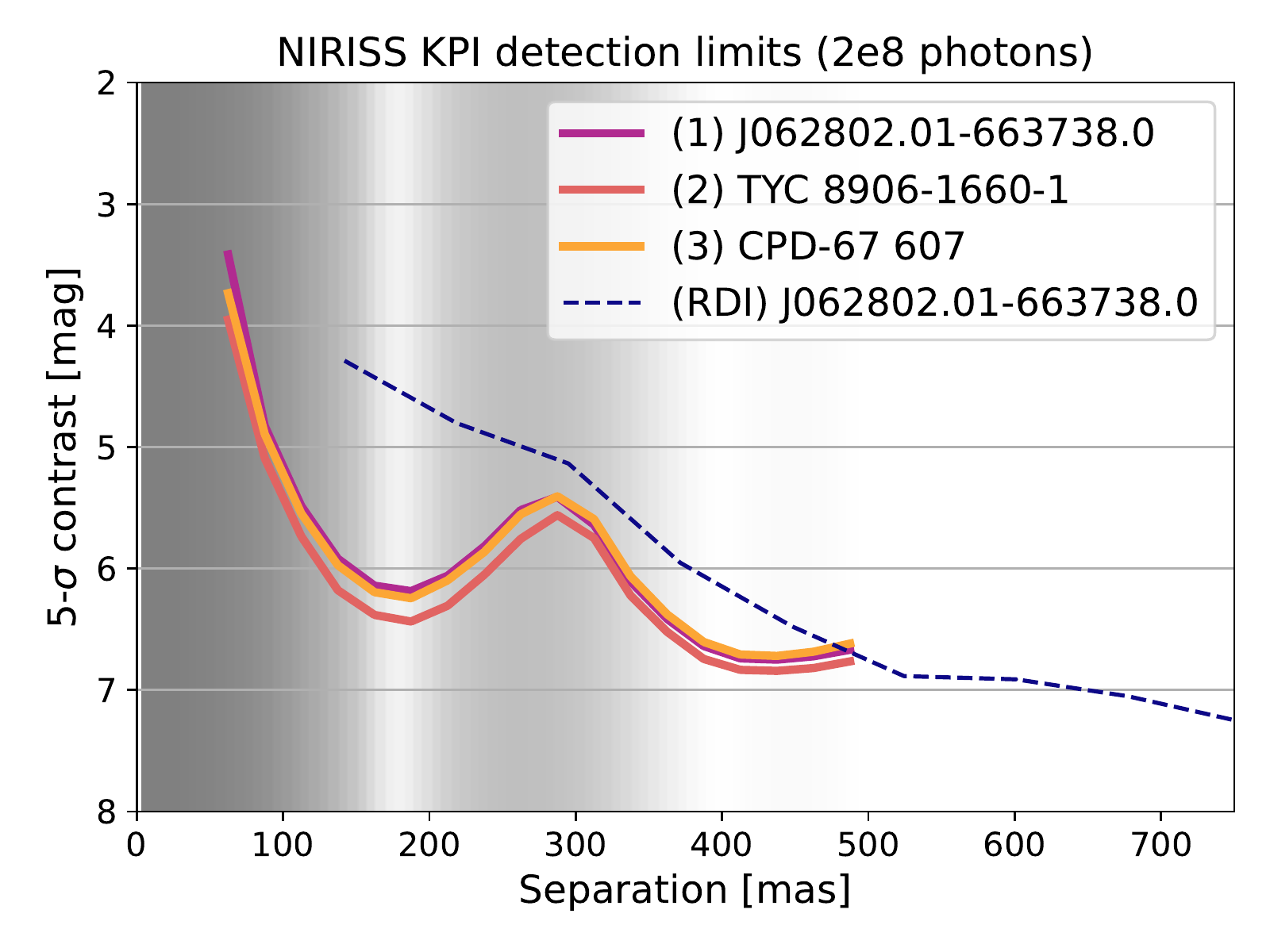}
\caption{Companion detection limits for the three point-source reference targets observed with KPI at $4.8~\text{\textmu m}$ (F480M) during \emph{JWST} NIRISS instrument commissioning. The observations were designed to collect $2\mathrm{e}{8}$ photons on the detector. The gray shaded background shows the azimuthal average of a NIRISS CLEARP F480M PSF computed with \texttt{WebbPSF} in a logarithmic color stretch where darker means brighter to highlight regions of increased photon noise. For comparison, the reference star differential (RDI) imaging contrast limits from the PSF subtraction with 2MASS~J062802.01-663738.0 are also shown. We note that this reduction suffers from imperfect image registration and the limits at small separations could likely still be improved.}
\label{fig:detlims}
\end{figure}

\begin{figure}[t!]
\centering
\includegraphics[width=\columnwidth]{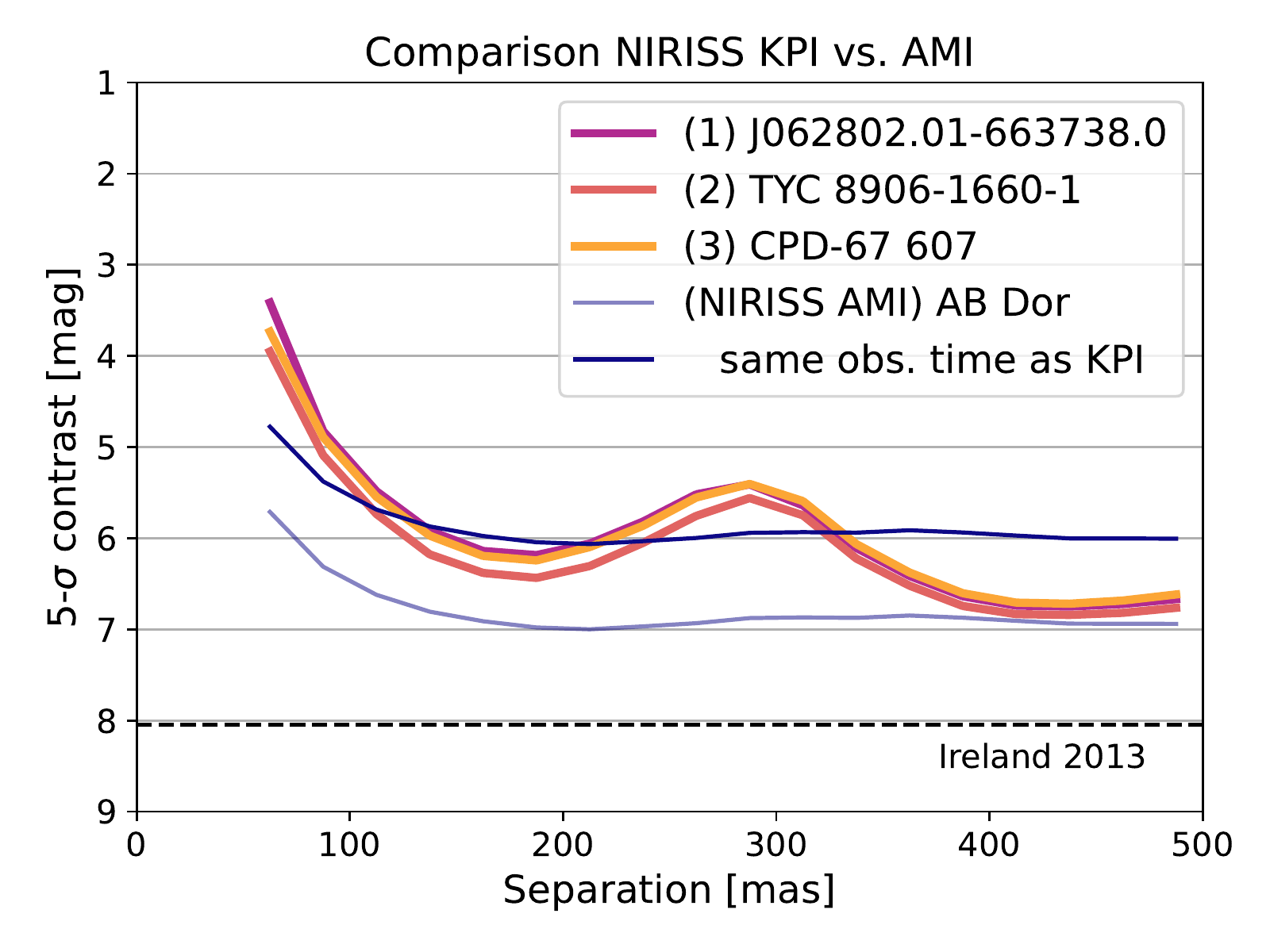}
\caption{Comparison between the NIRISS KPI and AMI companion detection limits measured during \emph{JWST} NIRISS instrument commissioning. Both the KPI and the AMI observations were designed to collect $2\mathrm{e}{8}$ photons on the detector and use the same F480M filter. The light blue curve shows the AMI detection limits extracted from the data and the dark blue curve shows the same limits after correcting them for the difference in throughput between the AMI and the KPI observations, i.e., multiplying them by $\sqrt{0.84/0.15}$, so that they correspond to the same amount of observing time. The fundamental photon noise floor for AMI according to \citet{ireland2013} is shown by a dashed black line.}
\label{fig:detlims_kpi_ami}
\end{figure}

Next, we compute companion detection limits for each of the three point-source reference targets from the second run. For each of the three targets, we use the other two targets as references and apply the Karhunen-Lo\`eve calibration in \texttt{fouriever} \citep{soummer2012,kammerer2019} with $K_\text{klip} = 2$ to calibrate out systematic errors. Then, we use the Absil method in \texttt{fouriever} to compute companion detection limits from the calibrated kernel phase observables of each target. For 2MASS~J062802.01-663738.0, we analytically remove the best fit companion before computing the detection limits. Figure~\ref{fig:detlims} shows the detection limits obtained from this procedure. The KPI detection limits show a prominent bump just inside of 300~mas separation which is caused by increased photon noise from the first Airy ring of the PSF. This is highlighted by the gray shaded background showing the azimuthal average of a NIRISS CLEARP F480M PSF computed with \texttt{WebbPSF}\footnote{\url{https://github.com/spacetelescope/webbpsf}} \citep{perrin2012} in a logarithmic color stretch. For the best target, the detection limits reach $\sim6.5$~mag at $\sim200$~mas and $\sim7$~mag at $\sim400$~mas. These limits agree well with the expectation from the kernel phase calibration residuals of $\sigma_\text{KP} \sim 0.002$~rad between TYC~8906-1660-1 and CPD-67~607 as shown in Figure~\ref{fig:raw_kerphase_run2} and translating into a contrast limit of $\sim6.75$~mag. Compared to the theoretical KPI detection limits predicted by \citet{ceau2019} using their binary test $T_B$ (which is similar to our detection criterion), our on-sky limits are $\sim1$~mag worse. This is consistent with the $\sim1$~mag difference between the theoretical AMI detection limits from \citet{ireland2013} and the limits measured on-sky (see Figure~\ref{fig:detlims_kpi_ami}). We expect that some fraction of this difference is caused by charge migration which has not been accounted for in the simulations and theoretical predictions. However, a more detailed analysis of the systematic errors is still ongoing.

For comparison, Figure~\ref{fig:detlims_kpi_ami} shows the detection limits achieved with the NIRISS AMI F480M commissioning observations of AB~Dor together with our KPI detection limits. Both the KPI and the AMI observations were designed to collect $2\mathrm{e}{8}$ photons on the detector and use the same F480M filter. The AMI data (observations 12 and 13 of PID 1093) were reduced with the \texttt{jwst} stage 1 and 2 pipelines and closure phases and visibility amplitudes were extracted from the cleaned interferograms using a custom version of \texttt{ImPlaneIA}\footnote{\url{https://github.com/anand0xff/ImPlaneIA}} \citep{greenbaum2015}. Next, the observables of AB~Dor were calibrated against those of the two observed reference targets (HD~37093, observations 15 and 16, and HD~36805, observations 18 and 19) using median subtraction. Then, the best fit parameters of the known close-in companion AB~Dor~C (separation $\sim326$~mas) were obtained with \texttt{fouriever} as discussed in Section~\ref{sec:companion_search} using both closure phases and visibility amplitudes in the fit. Finally, the best fit companion was analytically subtracted from the data before computing the detection limits using the Absil method in \texttt{fouriever}. The NIRISS AMI commissioning data reduction and analysis is described in more detail in \citet{sivaramakrishnan2022}.

Figure~\ref{fig:detlims_kpi_ami} shows that AMI (light blue curve) reaches deeper contrasts than KPI, especially at small angular separations $\lesssim400$~mas. This is the case since the NRM only collects photons on non-redundant baselines whereas the vast majority of the photons collected in KPI with the full pupil are affected by redundancy noise, so that the AMI observations achieve better detection limits than the KPI observations with the same number of collected photons. For a fair comparison, it needs to be considered that the NRM needs 5.6~times more time to collect the same number of photons than the full pupil since the NRM has an optical throughput of only 15\% while the CLEARP full pupil mask has an optical throughput of 84\%. Assuming that the contrast scales with the square root of the number of collected photons \citep[e.g.,][]{ireland2013} and multiplying the AMI detection limits by $\sqrt{0.84/0.15}$ results in more comparable limits between AMI (dark blue curve) and KPI. While the scaled AMI contrast curve still achieves better limits at $\sim250$--325~mas where KPI is suffering from increased photon noise from the first Airy ring of the PSF, KPI now achieves better limits beyond $\sim325$~mas where AMI is limited by the reduced throughput and uv-sampling. Finally, we note that the unscaled AMI detection limits (light blue curve) are still $\sim1$~mag above the fundamental noise floor according to \citet{ireland2013} and future efforts will aim at further improving the performance of NIRISS AMI and KPI \citep{sivaramakrishnan2022}.

\section{Summary \& conclusions}
\label{sec:summary_and_conclusions}

Fourier plane imaging techniques such as KPI and AMI are vital to explore close-in circumstellar environments with \emph{JWST} in order to directly detect substellar companions or study circumstellar dust \citep{artigau2014}. In this paper, we present the \texttt{Kpi3Pipeline} data reduction pipeline for kernel phase imaging with \emph{JWST} and the \texttt{fouriever} model fitting toolkit to search for companions and compute detection limits from KPI, AMI, and long-baseline interferometry data while accounting for correlated uncertainties in the fits. Then, we use these tools to reduce and analyze the \emph{JWST} NIRISS full pupil KPI data taken during the instrument commissioning. We also motivate and describe the new KPFITS file format for efficiently exchanging KPI and AMI data.

Among the four KPI targets observed during the instrument commissioning, we discover a low-contrast close-in companion candidate around CPD-66~562. Due to the low contrast (flux ratio of only $\sim20\%$), this companion candidate lies in the stellar and not the substellar regime. Moreover, with $2\mathrm{e}{8}$ photons collected on the detector at $4.8~\text{\textmu m}$, NIRISS KPI achieves 5--$\sigma$ companion detection limits of $\sim6.5$~mag at $\sim200$~mas and $\sim7$~mag at $\sim400$~mas. This is $\sim1$~mag worse than theoretical predictions, consistent with the $\sim1$~mag difference between theoretically predicted AMI detection limits and those measured on-sky. A detailed analysis of the kernel phase observables extracted from each of the four different dither positions that were tried during the instrument commissioning suggests that the first dither position of the second run yields an increased scatter in the raw and calibrated observables. Furthermore, we find a high-contrast ($\sim0.6\%$ flux ratio) companion at $\sim240$~mas separation around 2MASS~J062802.01-663738.0 using both kernel phase as well as classical PSF subtraction techniques. The host star is flagged as a giant in the TESS-HERMES survey \citep{sharma2018} so that the detected companion candidate (if not a background source) would also be in the stellar regime. A more detailed analysis of the detection and a comparison between Fourier plane and classical imaging techniques is planned for a future publication.

A comparison with NIRISS AMI commissioning observations of AB~Dor shows that when correcting for the different throughput between the NRM used for AMI and the CLEARP full pupil mask used for KPI, both techniques perform similar. We find that AMI achieves slightly deeper limits at the smallest separations of $\lesssim100$~mas and between $\sim250$--325~mas where KPI suffers from an increased photon noise from the core and the first Airy ring of the PSF, respectively. In the other regions, KPI performs slightly better due to its better uv-coverage and more compact PSF. However, we note that a major difference between NIRISS AMI and KPI is the range of targets they can observe. While AMI is well suited for bright and nearby targets (bright source limit of $m = 3$~mag in F480M) for which KPI might saturate quickly, KPI is beneficial for observing faint and more distant targets due to its 5.6 times higher throughput. In comparison with NIRCam KPI, NIRISS KPI offers a more precise target acquisition allowing to center the PSF on the exact same detector pixel every time which is not possible with NIRCam KPI. However, NIRCam KPI observations enable simultaneous collection of data in the short wavelength and the long wavelength channels, roughly doubling the observing efficiency. Compared to ground-based extreme adaptive optics-fed instruments, \emph{JWST} performs better for stars fainter than $\sim10$~mag in the L- and M-band. NIRISS AMI and KPI are expected to outperform current ground-based NRM and PSF subtraction methods at small separations within $\sim100/150$~mas in the L/M-band due to \emph{JWST}'s high thermal stability. In the future, however, METIS at the E-ELT is expected to reach contrasts of $\sim2\times10^{-5}$ at similar separations \citep{brandl2021}.

While \texttt{fouriever} is a stand-alone toolkit for performing calibrations, modeling correlations, and searching for companions in KPI, AMI, and long-baseline interferometry data, we specifically develop this toolkit to enable a uniform and state-of-the-art analysis of both \emph{JWST} AMI and KPI observations. In combination with the \texttt{Kpi3Pipeline} kernel phase stage 3 pipeline also introduced in this work, \texttt{fouriever} provides a powerful and easy-to-use solution for a kernel phase data reduction and analysis of \emph{JWST} full pupil images to the community. While the current version of the \texttt{Kpi3Pipeline} is compatible with NIRISS and NIRCam images, support for MIRI will be added in a future version.

Finally, we collect a number of recommendations mentioned throughout the paper for the convenience of future observers interested in using the KPI technique with \emph{JWST}:
\begin{itemize}
    \item KPI requires observations of a PSF reference target. To minimize calibration errors, this target should be of similar spectral type as the science target.
    \item Dithering is not recommended with NIRISS. The target acquisition accuracy and pointing stability of NIRISS are sufficient to repeatedly put a target on the same detector position. This is not necessarily true for other instruments and observing modes onboard \emph{JWST} which might be used for KPI in the future.
    \item While offsets can be used to place targets anywhere on the NIRISS detector, it is recommended to chose \textit{one} of the two predefined positions since these are located on a clean region close to the center of the detector. During commissioning, we found that the first dither position should be avoided since it yields a $\sim3$ times larger scatter in the raw kernel phase and a $\sim2$ times larger scatter in the kernel phase residuals.
    \item The \texttt{Kpi3Pipeline} introduced here extracts kernel phase observables from stage 2-reduced \emph{JWST} images. Before running the \texttt{Kpi3Pipeline}, users should reduce their data with the \texttt{jwst} stage 1 and 2 calibration pipelines while skipping the inter-pixel capacitance, the photometry, and the resample step.
    \item Users should run at least steps 1 (bad pixel cleaning), 2 (recentering), and 4 (kernel phase extraction) of the \texttt{Kpi3Pipeline} to ensure that they obtain scientifically useful data products (or perform their own custom bad pixel cleaning and recentering before extracting kernel phases).
\end{itemize}

\section{Acknowledgments}
\label{sec:acknowledgments}

These observations were made possible through the efforts of the many hundreds of people in the international commissioning team for JWST. This work is based on observations made with the NASA/ESA/CSA James Webb Space Telescope. The data were obtained from the Mikulski Archive for Space Telescopes at the Space Telescope Science Institute, which is operated by the Association of Universities for Research in Astronomy, Inc., under NASA contract NAS 5-03127 for JWST. These observations are associated with program \#1093. Support for programs \#1194, \#1411, and \#1412 was provided by NASA through a grant from the Space Telescope Science Institute, which is operated by the Association of Universities for Research in Astronomy, Inc., under NASA contract NAS 5-03127. This research has made use of the Spanish Virtual Observatory (\url{https://svo.cab.inta-csic.es}) project funded by MCIN/AEI/10.13039/501100011033/ through grant PID2020-112949GB-I00. F.M. acknowledges support from from the European Research Council (ERC) under the European Union’s Horizon 2020 research and innovation program (grant agreement CoG-683029). A.C. acknowledges support by the Heising-Simons Foundation through grant 2020-1825. A.G. acknowledges support from the ANID-ALMA fund No. ASTRO20-0059. D.J.~is supported by NRC Canada and by an NSERC Discovery Grant. J.S.B. acknowledges the full support from the CONACyT ``Ciencia de Frontera'' project CF-263975. M.R. would like to acknowledge funding from the Natural Sciences and Research Council of Canada (NSERC), as well as from the Fonds de Recherche du Qu\'ebec - Nature et Technologies (FRQNT) and the Institut de Recherche sur les Exoplan\`etes (iREx). T.V. acknowledges support from the Fonds the Recherche du Qu\'ebec - Nature et Technologies (FRQNT) and the Institut de Recherche sur les Exoplan\`etes (iREx). The manuscript was substantially improved following helpful comments from an anonymous referee.

\bibliographystyle{apj.bst}
\bibliography{bibliography.bib}

\begin{appendix}

\section{Comparison between fouriever and CANDID}
\label{sec:comparison_between_fouriever_and_candid}

\begin{figure*}[h!]
\centering
\includegraphics[width=\textwidth]{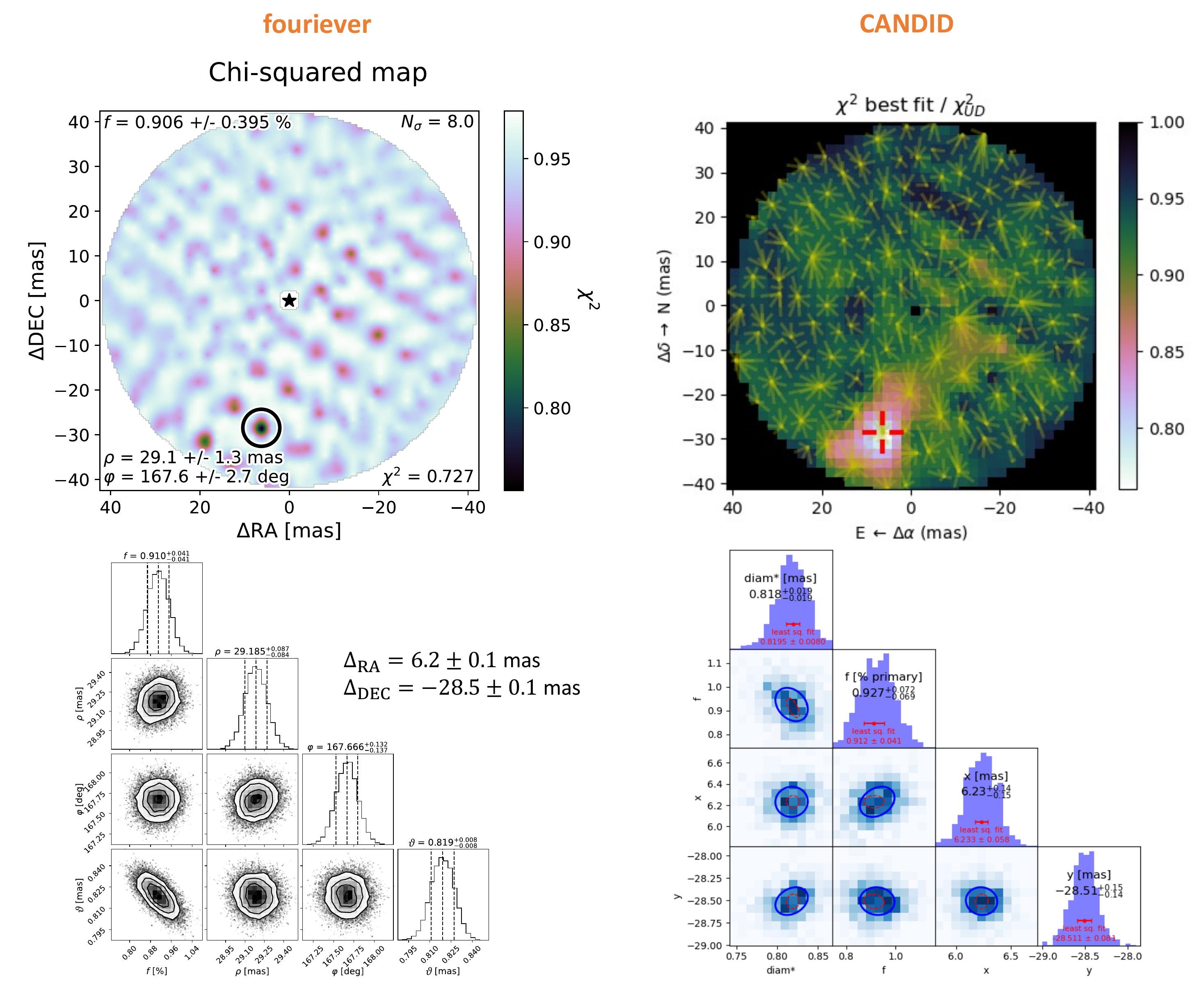}
\caption{Comparison of a companion search in VLTI/PIONIER data of AX~Cir with \texttt{fouriever} (left panels) and \texttt{CANDID} (right panels). While the $\chi^2$-detection map from \texttt{fouriever} (top left panel) shows the reduced $\chi^2$ of the binary model, the one from \texttt{CANDID} (top right panel) shows the ratio of the reduced $\chi^2$ of the binary model and the best fit uniform disk only model. The reduced $\chi^2$ of the best uniform disk only model is 0.975 in both cases. The best fit companion position is highlighted with a black circle and a red crosshair, respectively. The uncertainties in the top left panel are obtained from a least squares gradient descent minimization and are typically unreliable while the uncertainties in the bottom panels quote the 16th and 84th percentiles of the posterior distribution from the MCMC and the bootstrapping, respectively, and are more credible. Considering these uncertainties, \texttt{fouriever} and \texttt{CANDID} agree with each other to within one sigma.}
\label{fig:axcir_companion}
\end{figure*}

\begin{figure*}[h!]
\centering
\includegraphics[width=\textwidth]{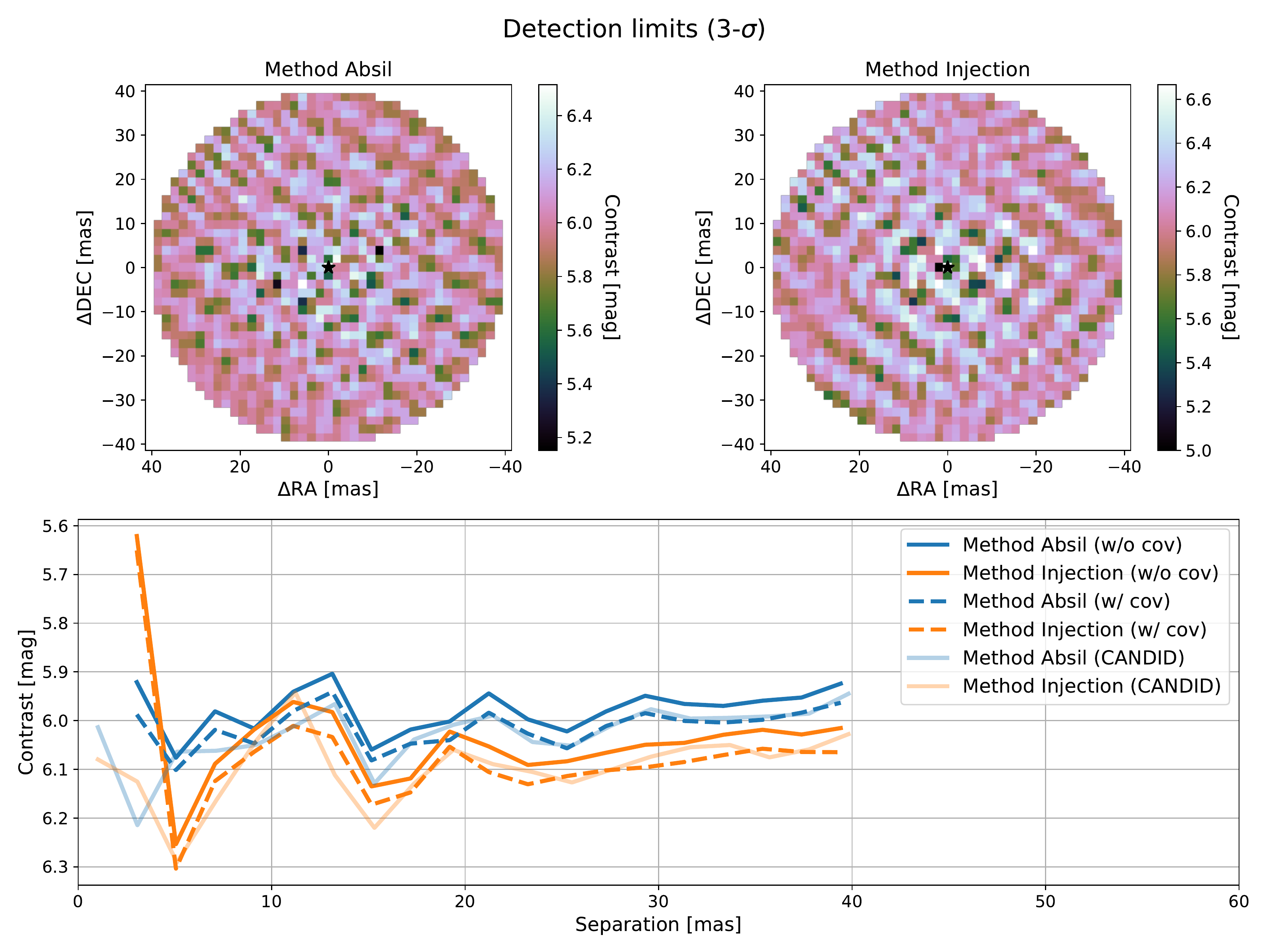}
\caption{Comparison of the 3--$\sigma$ companion detection limits obtained for VLTI/PIONIER data of AX~Cir with \texttt{fouriever} and \texttt{CANDID} after analytically removing the best fit companion shown in Figure~\ref{fig:axcir_companion}. The top left and right panels show the 2D detection limit maps obtained with the ``Absil'' and the ``Injection'' method using \texttt{fouriever} and the bottom panel shows the contrast curves (azimuthal averages) obtained from these maps with solid lines. The same two methods using \texttt{CANDID} yield the detection limits shown with solid transparent lines. The differences between \texttt{fouriever} and \texttt{CANDID} (especially at small angular separation) are caused by the detection limits being evaluated on slightly different grids (\texttt{fouriever} centers the host star in the center of a single pixel while \texttt{CANDID} centers it in between four pixels). The dashed lines show how the detection limits improve by $\sim5\%$ when accounting for basic correlations in the fits with \texttt{fouriever} (cf. Section~\ref{sec:detection_limits}).}
\label{fig:axcir_detlims}
\end{figure*}

\section{Kernel phase stage 3 pipeline parameters}
\label{sec:kernel_phase_stage_3_pipeline_parameters}

\begin{table*}[h!]
\caption{Tunable parameters of the \texttt{Kpi3Pipeline}. The FPNM, BCEN, COGI, and LDFT methods are described in more detail in the \texttt{XARA} documentation (see [1]).}
\centering
\begin{tabular}{lllll}
Name & Type & Default value & Allowed values & Description\\
\hline
\hline
\multicolumn{5}{c}{\textit{Bad pixel fixing step}}\\
\hline
\texttt{skip} & bool & \texttt{False} & \texttt{False}/\texttt{True} & Skip step?\\
\texttt{plot} & bool & \texttt{True} & \texttt{True}/\texttt{False} & Make diagnostic plot?\\
\texttt{bad\_bits} & list of str & \texttt{['DO\_NOT\_USE']} & See [2] & DQ flags considered bad pixels\\
\texttt{method} & str & \texttt{'medfilt'} & \texttt{'medfilt'}/\texttt{'fourier'} & Method to fix bad pixels\\
\hline
\multicolumn{5}{c}{\textit{Recentering step}}\\
\hline
\texttt{skip} & bool & \texttt{False} & \texttt{False}/\texttt{True} & Skip step?\\
\texttt{plot} & bool & \texttt{True} & \texttt{True}/\texttt{False} & Make diagnostic plot?\\
\texttt{method} & str & \texttt{'FPNM'} & \texttt{'FPNM'}/\texttt{'BCEN'}/\texttt{'COGI'} & Method to find PSF center\\
\texttt{trim} & bool & \texttt{True} & \texttt{True}/\texttt{False} & Trim images to square size?\\
\texttt{bmax} & float & \texttt{6.} & Positive non-zero & Maximum baseline [m] for FPNM\\
\texttt{pupil\_path} & str & \texttt{None} & Valid path & Path of custom pupil model\\
\hline
\multicolumn{5}{c}{\textit{Windowing step}}\\
\hline
\texttt{skip} & bool & \texttt{False} & \texttt{False}/\texttt{True} & Skip step?\\
\texttt{plot} & bool & \texttt{True} & \texttt{True}/\texttt{False} & Make diagnostic plot?\\
\texttt{wrad} & float & \texttt{None} & Positive non-zero & Windowing function radius [pix]\\
\hline
\multicolumn{5}{c}{\textit{Kernel phase extraction step}}\\
\hline
\texttt{skip} & bool & \texttt{False} & \texttt{False}/\texttt{True} & Skip step?\\
\texttt{plot} & bool & \texttt{True} & \texttt{True}/\texttt{False} & Make diagnostic plot?\\
\texttt{bmax} & float & \texttt{None} & Positive non-zero & Maximum baseline [m] for LDFT\\
\texttt{pupil\_path} & str & \texttt{None} & Valid path & Path of custom pupil model\\
\hline
\multicolumn{5}{c}{\textit{Empirical uncertainties step}}\\
\hline
\texttt{skip} & bool & \texttt{False} & \texttt{False}/\texttt{True} & Skip step?\\
\texttt{plot} & bool & \texttt{True} & \texttt{True}/\texttt{False} & Make diagnostic plot?\\
\texttt{get\_emp\_err} & bool & \texttt{True} & \texttt{True}/\texttt{False} & Use empirically est. errors?\\
\texttt{get\_emp\_cor} & bool & \texttt{False} & \texttt{False}/\texttt{True} & Use empirically est. correlations?\\
\hline
\multicolumn{5}{l}{\textbf{Notes.} \parbox[t]{15cm}{[1] = \url{https://github.com/fmartinache/xara},\\{[2]} = \url{https://jwst-reffiles.stsci.edu/source/data_quality.html}.}}
\end{tabular}
\label{tab:kernel_phase_stage_3_pipeline_parameters}
\end{table*}

\section{Analysis of dither positions}
\label{sec:analysis_of_dither_positions}

\begin{figure*}[h!]
\centering
\includegraphics[width=\textwidth]{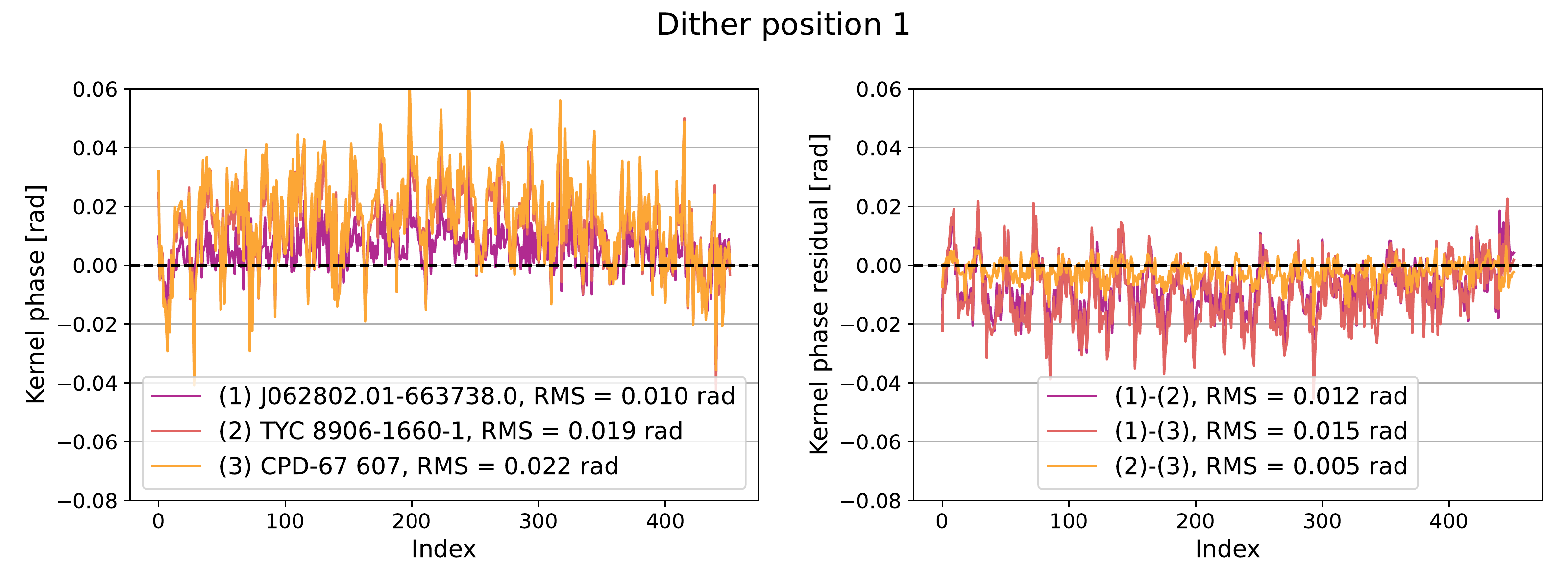}
\includegraphics[width=\textwidth]{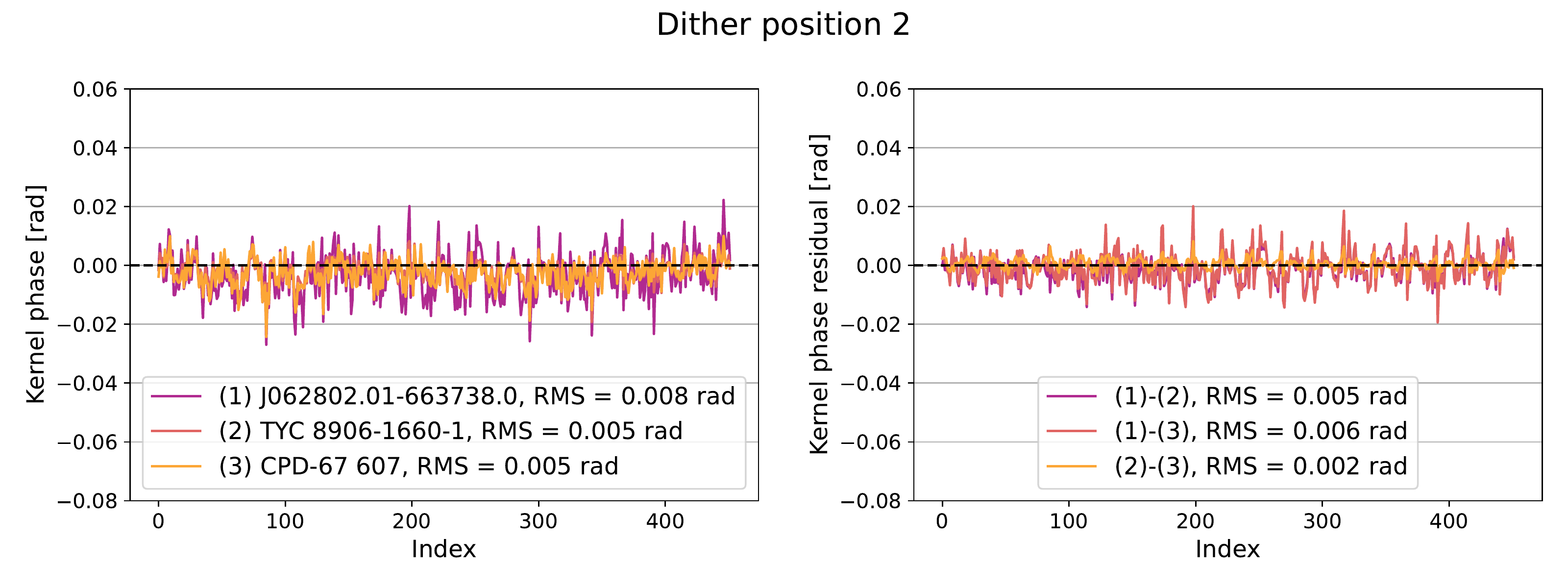}
\caption{Same as Figure~\ref{fig:raw_kerphase_run2}, but showing the raw kernel phase and the kernel phase residuals for each of the two dither positions from the second run separately.}
\label{fig:raw_kerphase_run2_dither}
\end{figure*}

\begin{figure*}[h!]
\centering
\includegraphics[width=\textwidth]{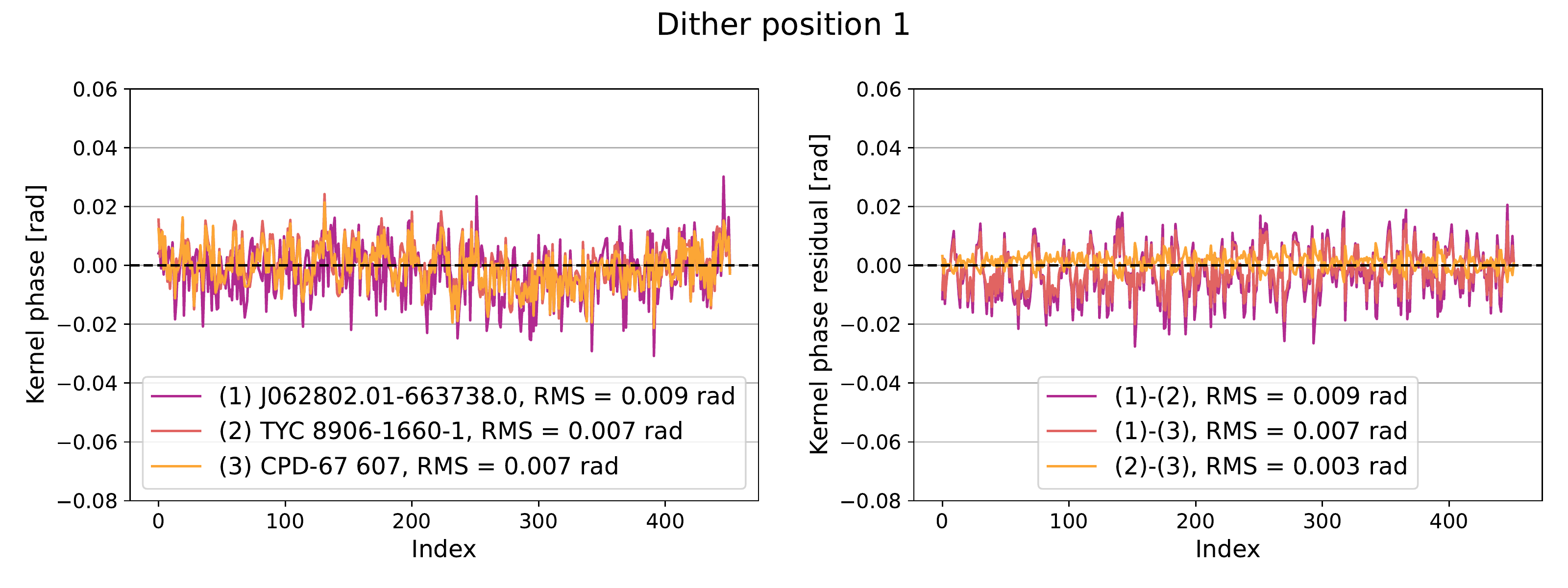}
\includegraphics[width=\textwidth]{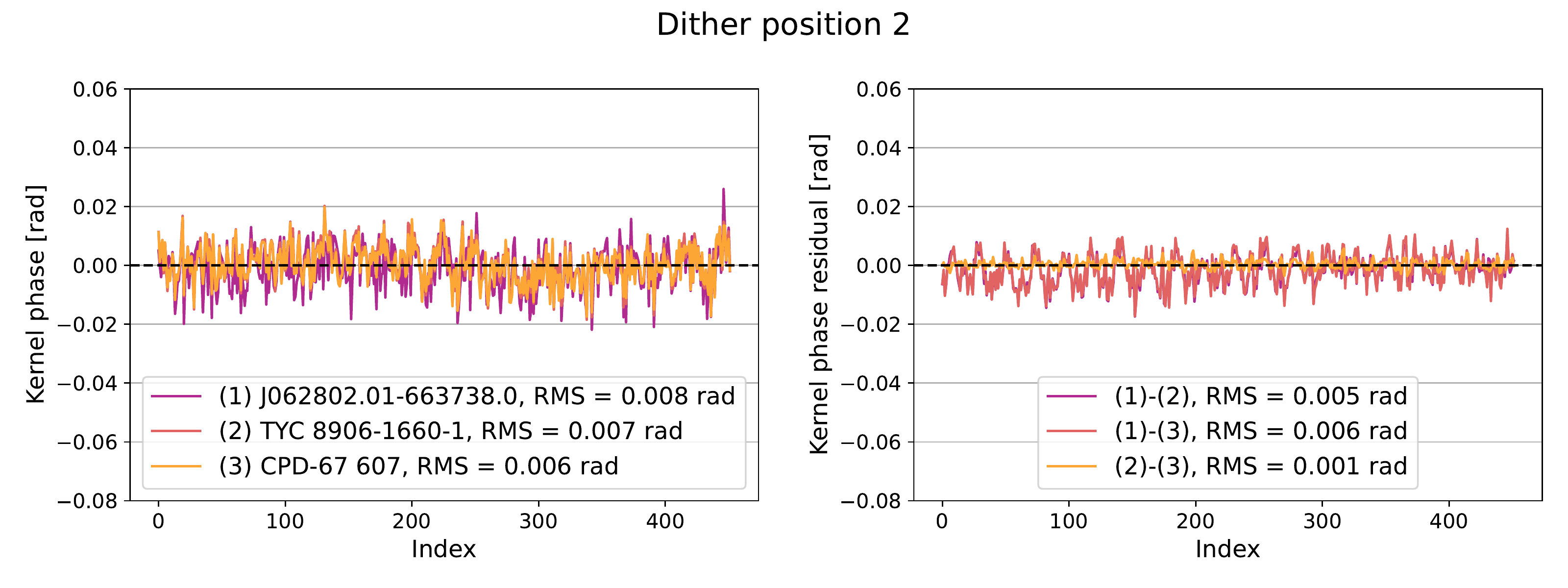}
\caption{Same as Figure~\ref{fig:raw_kerphase_run2}, but showing the raw kernel phase and the kernel phase residuals for each of the two dither positions from the first run separately.}
\label{fig:raw_kerphase_run1_dither}
\end{figure*}

\section{Analysis of outlier target (2MASS~J062802.01-663738.0)}
\label{sec:analysis_of_outlier_target}

\begin{figure*}[h!]
\centering
\includegraphics[width=0.49\textwidth]{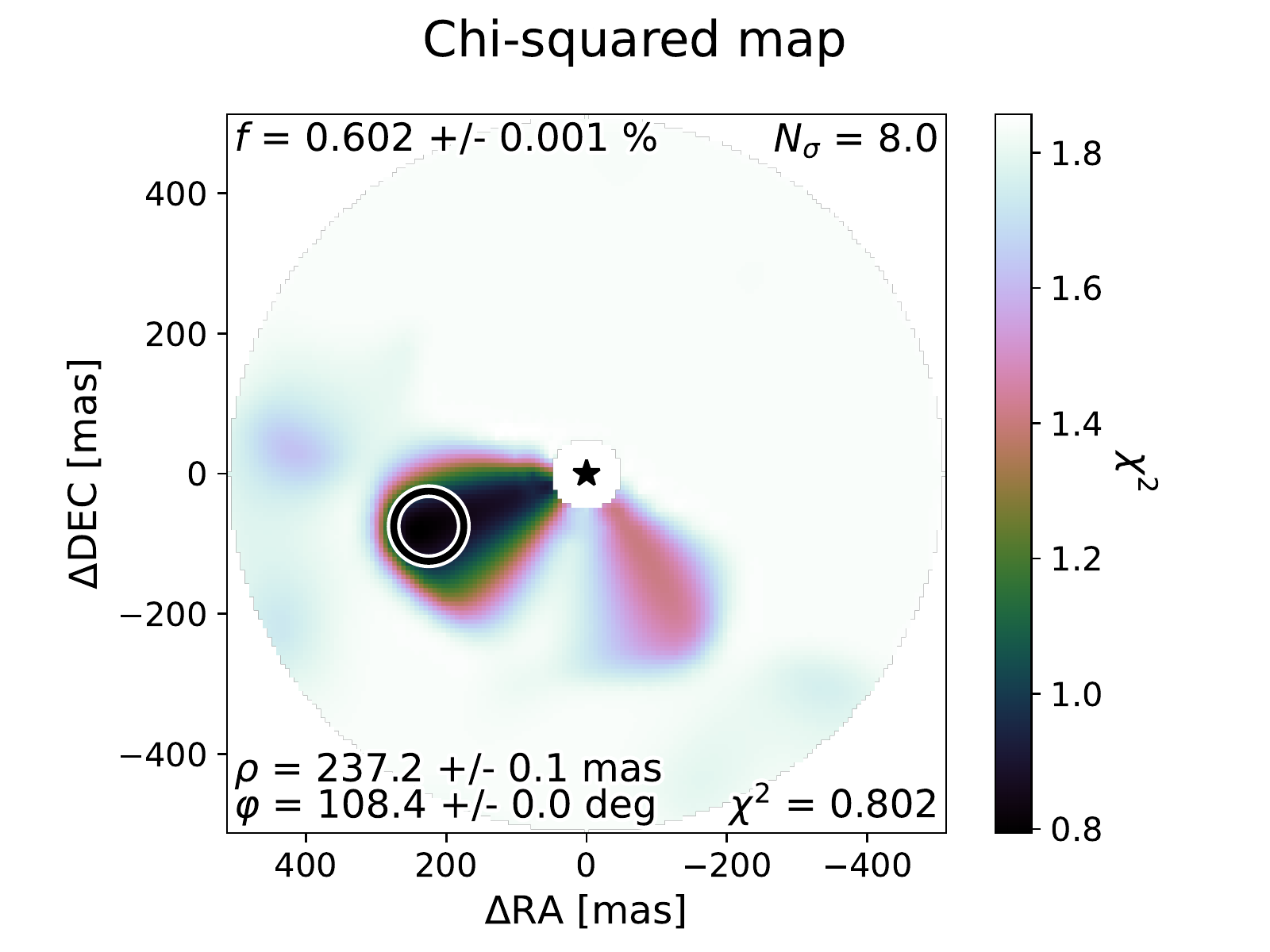}
\includegraphics[width=0.49\textwidth]{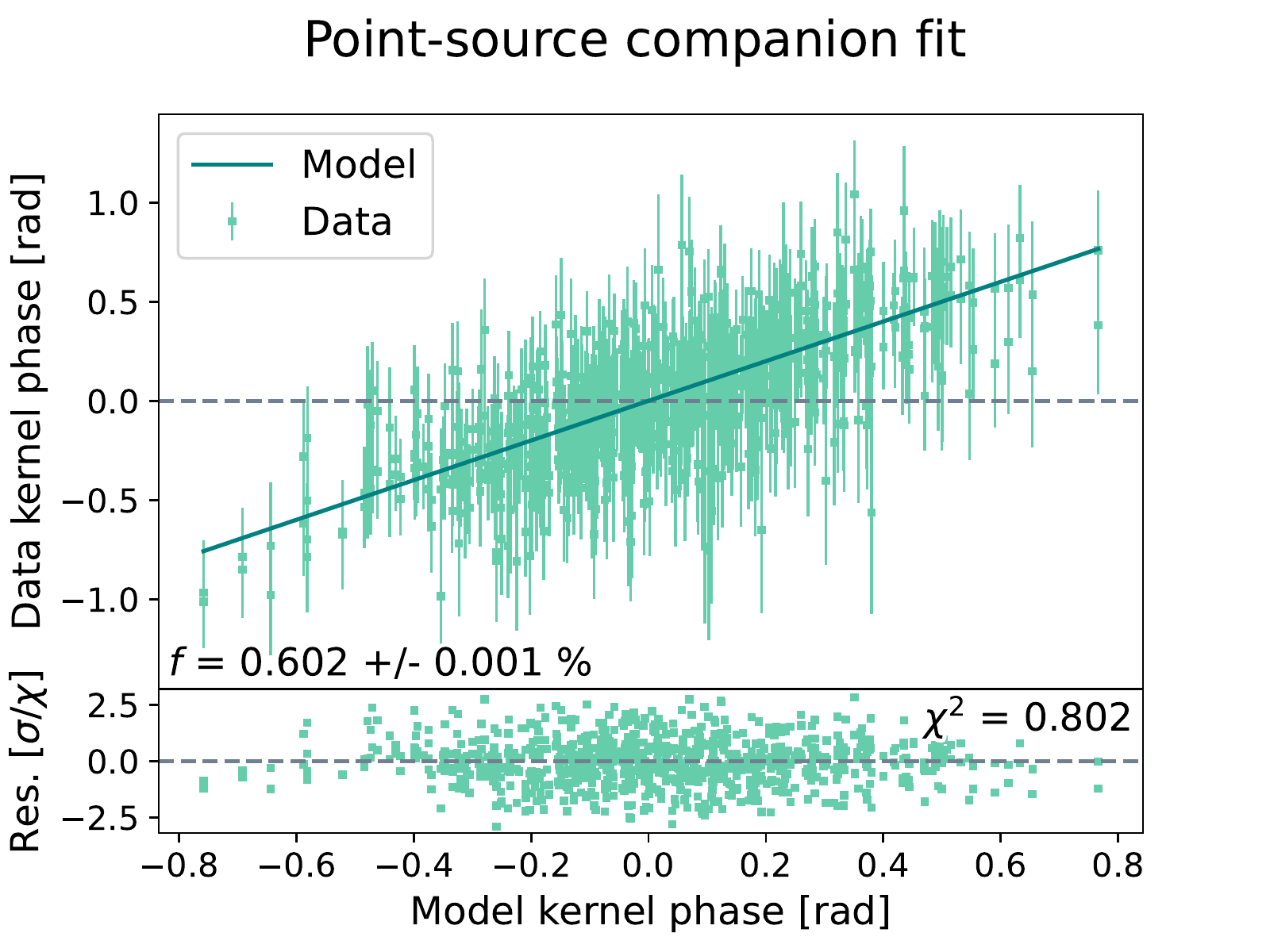}
\includegraphics[width=0.30\textwidth]{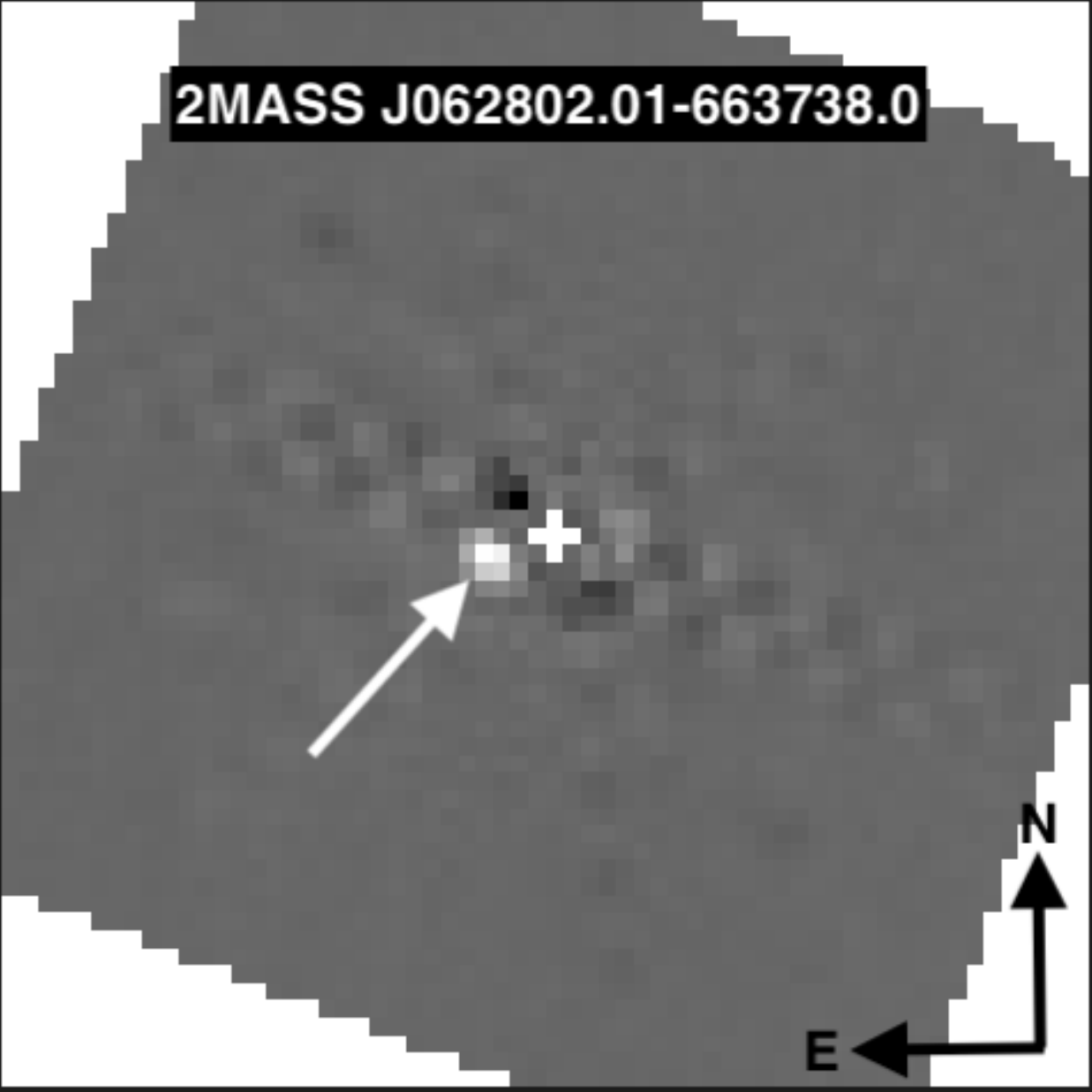}
\caption{Companion search around 2MASS~J062802.01-663738.0, the (supposedly) point-source reference target that was identified to be an outlier in both NIRISS KPI observing runs. The top left panel shows the $\chi^2$-detection map and the top right panel shows the data vs. model kernel phase signal as in Figure~\ref{fig:cpd-66_562_fouriever}, obtained from a binary model fit to the data from the second run. The bottom panel shows a PSF-subtracted image of 2MASS~J062802.01-663738.0 obtained with \texttt{pyKLIP} (20 KL modes). A companion candidate consistent with the one from the kernel phase reduction is clearly detected (white arrow). The image pixel scale is equivalent to the NIRISS detector pixel scale ($\sim65$~mas).}
\label{fig:outlier_companion_search}
\end{figure*}

\end{appendix}

\end{document}